\documentclass[12pt,a4paper,english]{JHEP3}
\usepackage[T1]{fontenc}
\usepackage[latin1]{inputenc}
\usepackage{graphicx}

\makeatletter

\usepackage{graphicx}

\usepackage{amsmath,cite,graphicx}
\title{{\textbf{Simulation of QED Radiation in Particle decays using the YFS Formalism}}}
\author{Keith Hamilton and Peter Richardson \\ Institute of Particle Physics Phenomenology, Department of Physics \\ University of Durham,  Durham, DH1 3LE, UK \\ Email: \email{keith.hamilton@durham.ac.uk, peter.richardson@durham.ac.uk}} \preprint{IPPP/06/14 \\DCPT/06/28}
\abstract{In this paper we describe a program ({\sf{SOPHTY}}) implementing QED
 corrections to decays in the {\sf{HERWIG++}} event generator. In order
 to resum the dominant soft emissions to all orders, the program is based
 on the YFS formalism.  In addition, universal large collinear logarithms
 are included and the approach can be systematically extended to incorporate
 exact, process specific, higher order corrections to decays.
 Due to the large number of possible decay modes the program is designed to 
 operate, as far as possible, independently of the details of the decay matrix
 elements.}

\newcommand{\newc}{\newcommand}
\newc{\gev}{\,GeV}
\newc{\mev}{\,MeV}
\newc{\ra}{\rightarrow}
\newc{\rpv}{$\mathrm{\not\!R_p}$}
\newc{\rp}{$\mathrm{R_p}$}
\newc{\real}{\mathcal{R}e}
\newc{\alsm}{{\displaystyle \sum_{\alpha=1,2}}}
\newc{\besm}{{\displaystyle \sum_{\beta=1,2}}}
\newc{\al}{\alpha}
\newc{\sgn}{\mr{sgn}\,}
\newc{\be}{\beta}
\newc{\ga}{\gamma}
\newc{\de}{\delta}
\newc{\sla}{\!\!\!\!\!\not\:\:\!}
\newc{\slab}{\!\!\!\!\!\not\,\,\,}
\newc{\slac}{\!\!\!\!\!\!\!\not\,\,\,\,}
\newc{\met}{$\not\!\!E_T$}
\newc{\cw}{\cos\theta_W}
\newc{\sw}{\sin\theta_W}
\newc{\ssw}{\sin^2\theta_W}
\newc{\ccw}{\cos^2\theta_W}
\newc{\cbe}{\cos\beta}
\newc{\sbe}{\sin\beta}
\newc{\ort}{\frac1{\sqrt{2}}}
\newc{\sh}{\hat{s}}
\newc{\uh}{\hat{u}}
\newc{\tha}{\hat{t}}
\newc{\sa}{\sin\al}
\newc{\ca}{\cos\al}
\newc{\mz}{M_{\mr{Z}}}
\newc{\mw}{M_{\mr{W}}}
\newc{\bv}{$\mathrm{\not\!B}$}
\newc{\lv}{$\mathrm{\not\!L}$}
\newc{\beq}{\begin{equation}}
\newc{\eeq}{\end{equation}}
\newc{\ie}{{\it i.e.\/}\ }
\newc{\lam}{\lambda}
\newc{\cht}{\tilde{\chi}}
\newc{\glt}{\tilde{g}}
\newc{\upt}{\tilde{u}}
\newc{\qkt}{\tilde{q}}
\newc{\elt}{\tilde{\ell}}
\newc{\hgt}{\tilde{H}}
\newc{\nut}{\tilde{\nu}}
\newc{\dnt}{\tilde{d}}
\newc{\ftl}{\mr{\tilde{f}}}
\newc{\psb}{\bar{\psi}}
\newc{\rtt}{\sqrt{2}}
\newc{\mut}{\tilde{\mu}}
\newc{\mr}{\mathrm}
\newc{\bath}{\bar{\theta}}
\newc{\tht}{\theta}
\newc{\JC}{{\bf J}}
\newc{\lra}{\longrightarrow}
\newc{\eg}{{\it e.g.\  }}
\newc{\barr}{\begin{eqnarray}}
\newc{\earr}{\end{eqnarray}}
\newc{\me}{\mathcal{M}}
\newc{\dbm}{\partial_\mu}
\newc{\dbmu}{\stackrel{\leftrightarrow\  }{\partial^\mu}}
\newc{\sgm}{\sigma_\mu}
\newc{\captionB}[2]{\caption[{#1}]{{\small {#2}}}}
\usepackage{babel}

\usepackage{babel}
\makeatother
\begin{document}

\section{Introduction}

The use of Monte Carlo event generators has become an essential part
of all experimental analyses, both in interpreting data from existing
experiments and in the design and planning of future experiments.
Given the crucial r\^{o}le which Monte Carlo simulations play in
experimental studies it is imperative that these simulations are as
accurate as possible.

While the existing Monte Carlo event generators have been highly successfully
over the last twenty years, it was realised that a new generation
of programs was necessary for the LHC. The reasons for this were twofold:
firstly a number of new ideas to improve the accuracy of the simulations
had been suggested \emph{e.g.} \cite{Richardson:2001df,Catani:2001cc,Frixione:2002ik,Gieseke:2003rz};
secondly the existing programs required major restructuring for long
term maintenance and to allow new theoretical developments to be incorporated.
Given the changing nature of computing in high energy physics, the
natural choice was to write these new programs in C++. A major effort
is therefore underway, in preparation for the LHC, to produce completely
new event generators~\cite{Gleisberg:2003xi}, as well as new versions
of established simulations~\cite{Bertini:2000uh,Lonnblad:2003wz,Gieseke:2003hm,Gieseke:2004af,Gieseke:2006rr}
in C++.

As part of the process of writing the new \textsf{HERWIG++} event
generator~\cite{Gieseke:2003hm,Gieseke:2004af,Gieseke:2006rr} we
wish to improve many aspects of the simulation process. One area where
improvements were needed was in the simulation of particle decays,
both of the fundamental particles produced in the perturbative stages
of the event and the decays of unstable hadrons. Several major improvements
have been made to the simulation of the decays: better modelling of
the matrix elements in hadronic decays; and full spin correlations
between the production and decay of particles~\cite{HwDecays}. Another
problem with the particle decays in the FORTRAN version of the \textsf{HERWIG}
program was the absence of QED radiation, which we will address in
this paper.

In existing Monte Carlo simulations the production of QED radiation
in particle decays is normally simulated using an interface to the
\textsf{PHOTOS} program~\cite{Barberio:1993qi,Barberio:1990ms,Golonka:2005pn}.
This program is based on the collinear approximation for the radiation
of photons together with corrections to reproduce the correct result
in the soft limit~\cite{Barberio:1993qi,Barberio:1990ms}. Recently,
it has been improved to include the full next-to-leading order QED
corrections for certain decay processes~\cite{Golonka:2005pn}.

Despite the success of \textsf{PHOTOS} it is based on the collinear
approximation for photon radiation. The production of radiation in
these decays is normally simulated in the rest frame of the decaying
particle. The kinematics of many of the decays, particularly of the
unstable hadrons, is such that the energy of the decay products is
not significantly larger than their mass, in which case we do not
expect the collinear limit to be a good approximation. However, there
is always a soft enhancement for the emission of QED radiation. We
therefore chose to base the simulation of QED radiation in \textsf{HERWIG++}
on the YFS~\cite{Yennie:1961ad} formalism for the resummation of
soft logarithms. This formalism has the major advantage that the exact
higher-order corrections can be systematically included, indeed the
majority of the most accurate simulations including higher-order QED
corrections are based on this approach \cite{JadachMPI,Bonneau-Martin,YFS2,Jadach:2000ir,Jadach:1999vf,Placzek:2003zg,Jadach:2001mp,Jadach:2001uu}. 

Another significant improvement arises from the use of the C++ programming
language and an object-oriented design for the program. The code framework
governing decays, in the \textsf{HERWIG++} program, is arranged so
that users can easily introduce the matrix element for a particular
decay mode using the C++ inheritance mechanism. This framework also
allows the inclusion of additional next-to-leading order matrix elements
for both standard \textsf{HERWIG++} and user defined decays. This
makes it possible for the leading-order matrix elements and their
higher-order corrections to be implemented in a systematic and consistent
manor, rather than relying on one program to handle the leading-order
decay and another for the higher-order corrections. This will be of
particular importance for the implementation of spin correlation effects.
Moreover, it will be easier for users to introduce new decays, including
higher-order corrections, as they will only require knowledge of,
and make modifications to, one program rather than two.

Note that, in full generality, it is not possible to consider radiative
corrections to production and decay processes separately%
\footnote{The case of $\mathrm{W}$ boson production is a noteworthy example
since it is both charged and unstable. This case has been studied
in detail in the context of the charged Drell-Yan process \cite{Baur:1995aa,Baur:2001ze,Baur:2004ig,CarloniCalame:2003ux,CarloniCalame:2005vc,Dittmaier:2001ay,Wackeroth:1996hz}
as well as single $\mathrm{W}$ production at LEP \cite{Argyres:1995ym,Beenakker:1996kn}. %
}, the minimal requirement for such a treatment to preserve gauge invariance
is that the intermediate particle be on-shell. For most applications
at hadron colliders we anticipate this to be a good approximation.
At a technical level this amounts to neglecting off-shell effects
in propagator numerators and including finite width effects in propagator
denominators via an overall Breit-Wigner factor that multiplies the
squared amplitude (the so-called \emph{narrow width} \emph{approximation}).
This approximation is already in effect at the level of the tree amplitudes
in the \textsf{HERWIG++} generator, in which spin correlation effects
are transmitted from the production stage to the decay stage according
to the algorithm described in \cite{Richardson:2001df}. As the \textsf{SOPHTY}
program is to dress tree-level events generated by the \textsf{HERWIG++}
simulation, a more subtle treatment including off-shell effects is
beyond the scope of this work. 

The same approximation scheme is adopted in many other generators
\emph{e.g.} the \emph{}\textsf{PHOTOS} and \textsf{WINHAC} \cite{Placzek:2003zg}
generators. Working in this approximation means that gauge invariance
of the QED corrections is then guaranteed by considering only the
universal leading log corrections to individualy decays in a cascade.
\emph{Exact} $\mathcal{O}\left(\alpha\right)$ corrections, comprising
of additional \emph{non-factorizable corrections} are process-specific,
and are therefore the subject of dedicated process-specific simulations;
the \textsf{SOPHTY} paradigm is primarily one of universality and
general applicability. Nevertheless, the dominant corrections are due to 
universal soft and collinear enhanced terms.

In the next section we will present our master equation, based on
the YFS formalism, for the generation of QED radiation, for the specific
case of particle decays. High multiplicity \emph{i.e.} greater than
two body, particle decays are normally simulated as a series of sequential
two body decays in \textsf{HERWIG++}. We therefore concentrate on
the cases of the decay of a neutral particle to two charged particles
and the decay of a charged particle to one charged and one neutral
particle. In addition, we present algorithms for the event generation,
using the master equation, for these two cases. The inclusion of higher-order
corrections to the decays is then considered in section~\ref{sec:Higher-Order-Corrections}
followed by a discussion of the results of the simulation. Finally
we present our conclusions and plans for further developments.

\section{Algorithms\label{sec:Master-Equation}}

We begin by considering the $n$-body decay of a particle in the \emph{absence}
of photonic radiation. The partial width for such a decay is given
by \begin{eqnarray}
\int\mathrm{d}\Gamma_{0} & = & \frac{1}{2M}\int\mathrm{d}\Phi_{q}\textrm{ }\left(2\pi\right)^{4}\delta^{4}\left(p-\sum_{i=1}^{n}q_{i}\right)\rho_{\alpha\beta}\mathcal{M}_{\alpha}\mathcal{M}_{\beta}^{*}\label{eq:2.1}\end{eqnarray}
 where\begin{eqnarray}
\mathrm{d}\Phi_{q} & = & \prod_{i=1}^{n}\frac{\mathrm{d}^{3}q_{i}}{\left(2\pi\right)^{3}2q_{i}^{0}},\label{eq:2.1b}\end{eqnarray}
 $M$ is the mass of the decaying particle, $p$ is its four-momentum
and $q_{i}$ is the momenta of the $i$th decay product. We have also
denoted the matrix element for the decay of a particle with helicity
$\alpha$ by $\mathcal{M}_{\alpha}$ and $\rho_{\alpha\beta}$ represents
the spin density matrix. In order to simplify the expressions we have
suppressed the dependence of the matrix element on the momenta of
the decaying particle and the decay products. In practice we do not
need to include the spin density matrix when calculating the total
width, if we wish to average over the helicities of the decaying particles
it is simply the identity matrix. However, inside the \textsf{HERWIG++}
simulation it allows us to include the correlation effects for the
decay.

In order to simulate the effects of additional QED radiation in the
decay, we must generalise (\ref{eq:2.1}) to include the effects of
arbitrary numbers of photons. In principle this extension is straightforward;
one simply replaces the matrix element and augments the phase space.
However, the matrix elements give rise to infrared divergences. The
cancellation of these soft divergences must be made explicit before
the Monte Carlo integration over the phase space can be performed.

In the YFS formalism this cancellation of the divergences is manifest
to all orders in perturbation theory. The cancellation relies on the
fact that in the divergent, soft photon, limit both real and virtual
corrections, to \emph{any} process, take the form of universal kinematic
factors multiplying the amplitude for that process without the additional
radiation. In summing over all of the different soft photon contributions,
these kinematic factors separately exponentiate, due to their universal
nature. The resulting product of exponentials is the manifestly finite
YFS form factor~\cite{Yennie:1961ad}. Residual, non-factorizing,
parts of the matrix elements, which cannot be exponentiated, are naturally
infrared finite.

Applying the YFS formalism to particle decays, by analogy to \cite{Ward:1987jg,Bonneau-Martin,YFS2},
we find that radiation modifies the $n$-body decay rate (\ref{eq:2.1})
to\begin{eqnarray}
\Gamma & = & \frac{1}{2M}\sum_{n_{\gamma}=0}^{\infty}\int\mathrm{d}\Phi\textrm{ }\frac{1}{n_{\gamma}!}e^{Y_{\textrm{total}}\left(\Omega\right)}\prod_{i=1}^{n_{\gamma}}\tilde{S}_{\textrm{total}}\left(k_{i}\right)\rho_{\alpha\beta}\mathcal{M}_{\alpha}\mathcal{M}_{\beta}^{*}\textrm{ }\mathcal{C}.\label{eq:2.2}\end{eqnarray}
This is the master equation from which we intend to generate the QED
radiation, where\begin{equation}
\mathrm{d}\Phi=\mathrm{d}\Phi_{p}\textrm{ }\mathrm{d}\Phi_{k}\textrm{ }\left(2\pi\right)^{4}\delta^{4}\left(p-P-K\right),\label{eq:2.2b}\end{equation}
 $K=\sum_{i=1}^{n_{\gamma}}k_{i}$ denotes the sum of photon momenta
and $P=\sum_{i=1}^{n}p_{i}$ denotes the sum of the primary decay
products momenta. The momentum of the $i$th decay product is $p_{i}$
(previously $q_{i}$ without the photon radiation), $k_{j}$ is the
momentum of the $j$th photon, while $\mathrm{d}\Phi_{p}$ and $\mathrm{d}\Phi_{k}$
are the associated phase space measures:\begin{eqnarray}
\mathrm{d}\Phi_{p} & = & \prod_{i=1}^{n}\frac{\mathrm{d}^{3}p_{i}}{\left(2\pi\right)^{3}2p_{i}^{0}};\label{eq:2.3}\\
\mathrm{d}\Phi_{k} & = & \prod_{i=1}^{n_{\gamma}}\frac{\mathrm{d}^{3}k_{i}}{k_{i}^{0}}\textrm{ }\Theta\left(k_{i},\Omega\right).\nonumber \end{eqnarray}
 The symbol $\Omega$ is used to denote the region of phase space
\emph{inside} which photons are soft and unresolvable, we choose this
to be the region $\left|\vec{k}\right|<\omega$, with $\omega$ an
energy cut-off. We define $\Theta\left(k_{i},\Omega\right)=0$ for
$k_{i}\in\Omega$ and $\Theta\left(k_{i},\Omega\right)=1$ for $k_{i}\not\in\Omega$.
This definition of $\Omega$ is not Lorentz invariant and in addition
to specifying a value for $\omega$ we must specify the frame in which
it has this value.

The total dipole radiation function \begin{equation}
\tilde{S}_{\textrm{total}}\left(k\right)=\sum_{i<j}^{n}\tilde{S}\left(p_{i},p_{j},k\right)\label{eq:2.4}\end{equation}
 is the sum of the individual dipole functions \begin{equation}
\tilde{S}\left(p_{i},p_{j},k\right)=\frac{\alpha}{4\pi^{2}}\textrm{ }Z_{i}\theta_{i}Z_{j}\theta_{j}\left(\frac{p_{i}}{p_{i}\cdot k}-\frac{p_{j}}{p_{j}\cdot k}\right)^{2},\label{eq:2.5}\end{equation}
 where $k$ is the four momentum of the photon, $Z_{i}$ is the charge
of the $i$th particle in units of the positron charge and $\theta_{i}=+1\left(-1\right)$
if the momentum $p_{i}$ is outgoing (incoming).

Likewise, the YFS form factor~\cite{Yennie:1961ad}, $Y_{\textrm{total}}\left(p_{i},p_{j},\Omega\right)$,
is given in terms of the form factor for pairs of charged particles
\begin{equation}
Y_{\textrm{total}}\left(p_{i},p_{j},\Omega\right)=\sum_{i<j}^{n}Y\left(p_{i},p_{j},\Omega\right).\label{eq:2.6}\end{equation}
 The YFS form factor for a pair of charged particles is given by \begin{equation}
Y\left(p_{i},p_{j},\Omega\right)=2\alpha\left(\mathcal{R}eB\left(p_{i},p_{j},\Omega\right)+\tilde{B}\left(p_{i},p_{j}\right)\right),\label{eq:2.7}\end{equation}
 where $\tilde{B}_{ij}$ is the integral of the dipole radiation function
over the soft photon phase space \begin{equation}
\tilde{B}\left(p_{i},p_{j},\Omega\right)=\frac{1}{8\pi^{2}}\textrm{ }Z_{i}\theta_{i}Z_{j}\theta_{j}\int_{\Omega}\frac{\mathrm{d}^{3}k}{\left|\vec{k}\right|}\left(\frac{p_{i}}{k\cdot p_{i}}-\frac{p_{j}}{k\cdot p_{j}}\right)^{2}.\label{eq:2.8}\end{equation}
 The virtual piece of the dipole \begin{equation}
B\left(p_{i},p_{j}\right)=-\frac{i}{8\pi^{3}}\textrm{ }Z_{i}\theta_{i}Z_{j}\theta_{j}\int\mathrm{d}^{4}k\textrm{ }\frac{1}{k^{2}}\left(\frac{2p_{i}\theta_{i}-k}{k^{2}-2k\cdot p_{i}\theta_{i}}+\frac{2p_{j}\theta_{j}+k}{k^{2}+2k\cdot p_{j}\theta_{j}}\right)^{2},\label{eq:2.9}\end{equation}
 does not depend on the cut-off, it is plainly Lorentz invariant.

The form factors are given for the case of a neutral particle decaying
to two charged final-state particles in Appendix \ref{sub:YFS_FF_Form_Factor}
and a charged particle decaying to one charged and one neutral particle
in Appendix~\ref{sub:YFS_IF_Form_Factor}, in the rest frame of the
decay products. For the case of a general moving frame only modifications
to the Lorentz variant bremsstrahlung integrals are needed, in the
final step where we have simplified with $\vec{p}_{i}=-\vec{p}_{j}$
and $E_{i}+E_{j}=M$. The corresponding $\tilde{B}_{ij}$ functions
are given in a general frame in \cite{Jadach:1995sp} and \cite{Placzek:2003zg}
respectively.

The factor $\mathcal{C}$ represents the total remainder of all of
the matrix elements contributing to the total decay width for the
particle, including any number of photons, when the leading soft divergent
pieces are exponentiated and cancelled. The contents of $\mathcal{C}$
are referred to as the \emph{infrared residuals}, they are infrared
finite and are written as a power expansion in the electromagnetic
coupling $\alpha$.

To order $\alpha$ there are three infrared residuals: the leading-order
matrix element $\left(\mathcal{O}\left(\alpha^{0}\right)\right)$,
the finite remainder of the one loop correction to the leading-order
process and the finite residual of the single photon emission matrix
element. Using superscripts to denote the order in $\alpha$ and subscripts
to denote the number of emitted photons, we have\begin{equation}
\mathcal{C}=1+\frac{1}{\rho_{\alpha\alpha'}\mathcal{M}_{\al}\mathcal{M}_{\al'}^{*}}\left(\bar{\beta}_{0}^{1}\left(p,\left\{ p_{i}\right\} \right)+\sum_{j=1}^{n_{\gamma}}\frac{\bar{\beta}_{1}^{1}\left(p,\left\{ p_{i}\right\} ,k_{j}\right)}{\tilde{S}_{\textrm{total}}\left(k_{j}\right)}+\mathcal{O}\left(\alpha^{2}\right)\right),\label{eq:2.10}\end{equation}
 where\begin{equation}
\begin{array}{lcl}
\bar{\beta}_{0}^{1}\left(p,\left\{ p_{i}\right\} \right) & = & \rho_{\alpha\beta}\left(\mathcal{M}_{\alpha}\mathcal{M}_{V\beta}^{1*}+\mathcal{M}_{V\alpha}^{1}\mathcal{M}_{\beta}^{*}-2\alpha B_{\textrm{total}}\mathcal{M}_{\alpha}\mathcal{M}_{\beta}^{*}\right)\\
\bar{\beta}_{1}^{1}\left(p,\left\{ p_{i}\right\} ,k\right) & = & \rho_{\alpha\beta}\left(\frac{1}{2\left(2\pi\right)^{3}}\mathcal{M}_{R\alpha}^{1}\mathcal{M}_{R\beta}^{*1}-\tilde{S}_{\textrm{total}}\left(k_{j}\right)\mathcal{M}_{\alpha}\mathcal{M}_{\beta}^{*}\right)\end{array}.\label{eq:2.11}\end{equation}
 In (\ref{eq:2.11}) we use $\mathcal{M}_{R}^{1}$ and $\mathcal{M}_{V}^{1}$
to denote the $\mathcal{O}\left(\alpha\right)$ corrections to the
leading-order matrix element $\left(\mathcal{M}\right)$ from single
real and single virtual photon corrections. The extension of the master
formula to higher orders in the infrared residuals $\left(\bar{\beta}\right)$
is straightforward, it is only limited by the usual technical difficulties
associated with calculating Feynman diagrams that involve many loops
or legs.

In practice the leading-order matrix element is only strictly defined
for the $n$-body phase space, not the phase space with additional
photons. We therefore need a procedure which maps the momenta of the
decay products after radiation, including the photons, on to the momentum
configuration of the decay products prior to the generation of any
radiation. Since the momenta of the primary decay products are to
be generated first before any QED radiation, according to the leading-order
distribution, we can use these momenta to calculate the leading-order
matrix element.

In order to implement the theoretical framework of the master equation
in an event generator, one expects that a number of different algorithms
must be devised to cope with all possible decay multiplicities and
charge configurations. In practice, due to the way in which particle
decays are simulated in \textsf{HERWIG++}, most of the decays will
either involve the decay of a charged particle to one charged and
one neutral particle, or the decay of a neutral particle to two charged
particles. Furthermore, we anticipate that more complicated decays
will proceed via repeated applications and simple extensions of these
two types of decay. We therefore concentrate on these cases.

\subsection{Final-Final Dipole\label{sub:Final-Final-Dipoles}}

The purpose of this algorithm is to \emph{dress} the decays, generated
by the core \textsf{HERWIG++} program, in which a neutral particle
decays to two charged particles, with QED radiation. The input to
our algorithm therefore consists of the momenta and quantum numbers
of the parent particle and its children, distributed according to
the leading-order differential decay rate (\ref{eq:2.1}).

In addition to infrared singularities, the dipole functions (\ref{eq:2.5})
also exhibit mass singular behaviour associated with small-angle photon
emission from the charged particles, in the massless limit. Therefore
the angles of the radiated photons with respect to the dipole must
remain fixed throughout the event generation process in order to produce
an efficient, stable algorithm.

In order to achieve this, we define the rest frame of the primary
decay products and generate the photons in this frame. In this respect
our approach is similar to that used in the $\mathcal{KK}$ event
generator \cite{Jadach:1999vf} for final-state radiation. Initially
the photons are generated according to the dipole functions, which
have a simple form in this frame. Implicit in the definition of the
rest frame of the primary decay products (in this case the dipole)
is the fact that the \emph{incoming} three-momentum of the decaying
particle must be equal to the \emph{total outgoing} three-momentum
of the photons. The three-momenta of the original decay products are
then rescaled (reduced) to ensure energy and momentum conservation.

A na\"{\i}ve application of the method outlined above will lead to
spurious results. It is important that we take into account the effects
of the aforesaid choice of frame on the phase-space integration measure.
To do this we employ the method of integration over the Lorentz group,
as described in \cite{torino}, to transform the phase-space measure
in (\ref{eq:2.2}). We start, by introducing the definition of the
momentum of the decaying particle and the total momentum of the primary
decay products in terms of $\delta$-functions, assuming the full
phase space is to be integrated over \emph{i.e}. \begin{eqnarray}
\int{\textrm{d}}\Phi\textrm{ } & = & \left(2\pi\right)^{4}\int\mathrm{d}\Phi_{p}\textrm{ }\mathrm{d}\Phi_{k}\textrm{ }\mathrm{d}^{4}p\textrm{ }\mathrm{d}s\textrm{ }\mathrm{d}^{4}P\textrm{ }2M\delta^{3}\left(p\right)\delta\left(p^{2}-M^{2}\right)\label{eq:3.1.1}\\
 &  & \delta^{4}\left(p-P-K\right)\delta^{4}\left(P-\sum_{i=1}^{n}p_{i}\right)\delta\left(P^{2}-s\right).\nonumber \end{eqnarray}
 Secondly we insert the identity \begin{equation}
\int\mathrm{d}^{4}X\textrm{ }\frac{2}{s^{2}}\delta\left(\frac{X^{2}}{s}-1\right)\delta^{3}\left(L^{-1}\frac{P}{\sqrt{s}}\right)=1,\label{eq:3.1.2}\end{equation}
 where $L^{-1}$ is the boost \emph{from} the frame in which $X=\left(X_{0},\vec{X}\right)$
\emph{to} the rest frame of $X$. This identity, in conjunction with
those already present in (\ref{eq:3.1.1}), constrains the boost $L$
to be the Lorentz transformation from the rest frame of the primary
decay products $\left(P\right)$ to the rest frame of the decaying
particle $\left(p\right)$. We then change the integration variables,
by applying the Lorentz boost $L$ to all of the momenta involved,
which is trivial as most of our expression (\ref{eq:3.1.1}) is Lorentz
invariant. This gives \begin{eqnarray}
\int\mathrm{d}\Phi & = & \left(2\pi\right)^{4}\int\mathrm{d}\Phi_{p}\textrm{ }\mathrm{d}\Phi_{k}\textrm{ }\mathrm{d}^{4}p\textrm{ }\mathrm{d}s\textrm{ }\mathrm{d}^{4}P\textrm{ }\mathrm{d}^{4}X\textrm{ }\delta^{3}\left(Lp\right)\delta\left(p^{2}-M^{2}\right)\label{eq:3.1.3}\\
 &  & \delta^{4}\left(p-P-K\right)\delta^{4}\left(P-\sum_{i=1}^{n}p_{i}\right)\delta\left(P^{2}-s\right)\textrm{ }\frac{4M}{s^{2}}\delta\left(\frac{X^{2}}{s}-1\right)\delta^{3}\left(\frac{P}{\sqrt{s}}\right).\nonumber \end{eqnarray}
 Integrating over the four momentum $P$, $X$ and $p$ we obtain
\begin{equation}
\int\mathrm{d}\Phi=\left(2\pi\right)^{4}\int\mathrm{d}\Phi_{p}\textrm{ }\mathrm{d}\Phi_{k}\textrm{ }\mathrm{d}s\textrm{ }\frac{s}{M^{2}}\delta\left(M^{2}-\left(P+K\right)^{2}\right)\delta^{4}\left(P-\sum_{i=1}^{n}p_{i}\right).\label{eq:3.1.4}\end{equation}
 The integral over $s$ can then be performed giving,\begin{equation}
\int\mathrm{d}\Phi=\int\mathrm{d}\Phi_{p}\textrm{ }\mathrm{d}\Phi_{k}\textrm{ }\frac{s}{M^{2}\left(1+\frac{K_{0}}{\sqrt{s}}\right)}\left(2\pi\right)^{4}\delta^{4}\left(P-\sum_{i=1}^{n}p_{i}\right).\label{eq:3.1.5}\end{equation}

As we first generate the momenta of the other decay products according
to the leading-order matrix element we need to rewrite the integral
in terms of the leading-order phase space. This is achieved by rescaling
the momenta of the decay products before radiation to give the correct
invariant mass for the decay products after the photon radiation.
We define a momentum rescaling factor, $u$, such that the three momenta
obey $\vec{q}_{i}=u\vec{p}_{i}$. The momentum rescaling $u$ is determined,
by on-shell constraints, to be the solution of \begin{equation}
\sum_{i}^{n}\sqrt{u^{2}\left|\vec{q}_{i}\right|^{2}+m_{i}^{2}}-\sqrt{s}=0,\label{eq:3.1.8}\end{equation}
 where $m_{i}^{2}=p_{i}^{2}=q_{i}^{2}$.

The unitary algorithm techniques of~\cite{Kleiss:1991rn} can be
used to show that \begin{equation}
\int\mathrm{d}\Phi_{p}\textrm{ }\delta^{4}\left(P-\sum_{i=1}^{n}p_{i}\right)=\int\mathrm{d}\Phi_{q}\delta^{4}\left(Q-\sum_{i=1}^{n}q_{i}\right)\frac{1}{u^{3}}\frac{\left(M-\sum_{i=1}^{n}\frac{m_{i}^{2}}{q_{i}^{0}}\right)}{\left(\sqrt{s}-\sum_{i=1}^{n}\frac{m_{i}^{2}}{p_{i}^{0}}\right)}\prod_{i=1}^{n}\frac{q_{i}^{0}u^{3}}{p_{i}^{0}}.\label{eq:3.1.9}\end{equation}
 With this result we can rewrite our phase space measure as {\small \begin{equation}
\int\mathrm{d}\Phi=\int\mathrm{d}\Phi_{q}\textrm{ }\mathrm{d}\Phi_{k}\left(2\pi\right)^{4}\delta^{4}\left(Q-\sum_{i=1}^{n}q_{i}\right)\frac{s^{\frac{3}{2}}}{M^{2}\left(\sqrt{s}+K_{0}\right)}\frac{1}{u^{3}}\frac{\left(M-\sum_{i=1}^{n}\frac{m_{i}^{2}}{q_{i}^{0}}\right)}{\left(\sqrt{s}-\sum_{i=1}^{n}\frac{m_{i}^{2}}{p_{i}^{0}}\right)}\prod_{i=1}^{n}\frac{q_{i}^{0}u^{3}}{p_{i}^{0}}.\label{eq:3.1.10}\end{equation}
}The decay width becomes {\small \begin{equation}
\Gamma=\sum_{n_{\gamma}=0}^{\infty}\frac{1}{n_{\gamma}!}\int\mathrm{d}\Gamma_{0}\mathrm{d}\Phi_{k}\textrm{ }e^{Y_{\textrm{total}}\left(\Omega\right)}\prod_{i=1}^{n_{\gamma}}\tilde{S}_{\textrm{total}}\left(k_{i}\right)\mathcal{C}\frac{s^{\frac{3}{2}}}{M^{2}u^{3}\left(\sqrt{s}+K_{0}\right)}\frac{\left(M-\sum_{i=1}^{n}\frac{m_{i}^{2}}{q_{i}^{0}}\right)}{\left(\sqrt{s}-\sum_{i=1}^{n}\frac{m_{i}^{2}}{p_{i}^{0}}\right)}\prod_{i=1}^{n}\frac{q_{i}^{0}u^{3}}{p_{i}^{0}}\label{eq:3.1.11}\end{equation}
}with $\mathrm{d}\Gamma_{0}$ given by (\ref{eq:2.1}). Equation (\ref{eq:3.1.11})
allows the construction of an algorithm in which the leading-order
subprocess may be generated independently of and prior to QED radiation.

Thus far we have treated the general case (an $n$ body final state)
but we will now specialise to the case of a neutral particle decaying
to two charged particles. In this case the rescaling factor is $u=|\vec{p}|/|\vec{q}|$
where $|\vec{p}|$ is the magnitude of the momentum of the decay products
in their rest frame after the radiation and $|\vec{q}|$ is the magnitude
of the momentum of the decay products in their rest frame before the
radiation. In this case the total width is \begin{equation}
\Gamma=\sum_{n_{\gamma}=0}^{\infty}\int\mathrm{d}\Gamma_{0}\textrm{ }\mathrm{d}\Phi_{k}\textrm{ }\frac{1}{n_{\gamma}!}e^{Y_{\textrm{total}}\left(\Omega\right)}\prod_{i=1}^{n_{\gamma}}\tilde{S}_{\textrm{total}}\left(k_{i}\right)\mathcal{C}\textrm{ }\frac{\sqrt{s}\left|\vec{p}\right|}{M\left|\vec{q}\right|}\left(1+\frac{K_{0}}{\sqrt{s}}\right)^{-1},\label{eq:3.1.7}\end{equation}
 where \[
\mathrm{d}\Gamma_{0}=\frac{1}{2M}\textrm{ }\mathrm{d}\Omega_{q_{2}}\frac{\left|\vec{q}\right|}{M}\rho_{\alpha\beta}\mathcal{M}_{\alpha}\mathcal{M}_{\beta}^{*}.\]

Up to now we have not made \emph{any} approximations other than the
truncation of the infrared residuals at $\mathcal{O}\left(\alpha\right)$.
To simulate events using these results, (\ref{eq:3.1.7}) and (\ref{eq:3.1.11}),
we need to make some approximations in order to obtain a distribution
which is fast and efficient to generate by Monte Carlo methods. It
is important to note that these simplifications are later \emph{exactly}
compensated by appropriate weighting and rejection of events. Naturally
the first simplification we make is to omit the higher-order, infrared
finite corrections represented by the factor $\mathcal{C}$. We also
neglect the factors associated with the rescaling of the leading-order
phase space. Both factors tend to one in the limit of soft QED radiation
and so neglecting them is reasonable, given that the vast majority
of photons produced will be soft. In addition to these two omissions
we also approximate the momenta $p_{1}$ and $p_{2}$ by the values
they would have in the absence of any QED radiation, $q_{1}$ and
$q_{2}$, this approximation is justified on the same grounds. These
simplifications give the following \textit{crude} distribution \begin{equation}
\Gamma_{\textrm{crude}}=\sum_{n_{\gamma}=0}^{\infty}\int\mathrm{d}\Gamma_{0}\mathrm{d}\Phi_{k}\textrm{ }\frac{1}{n_{\gamma}!}e^{Y\left(q_{1},q_{2},\Omega\right)}\prod_{i=1}^{n_{\gamma}}\tilde{S}\left(q_{1},q_{2},k_{i}\right).\label{eq:3.1.12}\end{equation}

Since we are working in the dipole rest frame, $\vec{q}_{1}=-\vec{q}_{2}$,
the kinematics simplify to the extent that the dipole function is
analytically integrable. Moreover it means the only dependence of
the QED part of (\ref{eq:3.1.12}) on $q_{1}$ and $q_{2}$ is through
their masses. Consequently we have the desired factorization that
(\ref{eq:3.1.12}) is really a product of two separate integrals,
one for the leading-order decay and one for the QED radiation. The
distributions may therefore be generated independently. Defining \begin{equation}
\bar{n}=\int\textrm{ }\frac{\mathrm{d}^{3}k_{i}}{k_{i}^{0}}\textrm{ }\Theta\left(k_{i},\Omega\right)\textrm{ }\tilde{S}\left(p_{1},p_{2},k_{i}\right),\label{eq:3.1.13}\end{equation}
 we obtain \begin{equation}
\Gamma_{\textrm{crude}}=\Gamma_{0}\sum_{n_{\gamma}=0}^{\infty}\frac{1}{n_{\gamma}!}\bar{n}^{n_{\gamma}}e^{-\bar{n}}=\Gamma_{0}.\label{eq:3.1.14}\end{equation}
 In deriving (\ref{eq:3.1.14}) we have also made the approximation
that the YFS form factor is $Y\approx-\bar{n}$. According to $\Gamma_{\mathrm{crude}}$
the photon multiplicity follows a Poisson distribution with average~$\bar{n}$.
In practice it is sometimes useful to neglect part of the dipole distribution
as described in Appendix~\ref{sub:Final-Final_Generation}.

Once such a decay has been generated in the main \textsf{HERWIG++}
code it may dressed with QED radiation using the following algorithm: 

\begin{enumerate}
\item The number of photons is generated according to a Poisson distribution
with average $\bar{n}$. 
\item The momenta of the photons is then generated as described in Appendix~\ref{sub:Final-Final_Generation}.
This gives the crude distribution. 
\item The exact distribution (the master equation for $\Gamma$) is obtained
using rejection techniques. The weight for rejection is given by \begin{equation}
\mathcal{W}=\mathcal{W}_{\textrm{dipole}}\times\mathcal{W}_{\textrm{YFS}}\times\mathcal{W}_{\textrm{Jacobian}}\times\mathcal{W}_{\textrm{higher}},\label{eq:3.1.15}\end{equation}
 where \begin{equation}
\begin{array}{lcl}
\mathcal{W_{\textrm{dipole}}} & = & \prod_{i=1}^{n_{\gamma}}\frac{\tilde{S}\left(p_{1},p_{2},k_{i}\right)}{\tilde{S}\left(q_{1},q_{2},k_{i}\right)},\\
\mathcal{W_{\textrm{YFS}}} & = & \frac{e^{Y\left(p_{1},p_{2},\Omega_{B}\right)}}{e^{-\bar{n}}},\\
\mathcal{W_{\textrm{Jacobian}}} & = & \frac{\sqrt{s}\left|\vec{p}\right|}{M\left|\vec{q}\right|}\left(\sqrt{s}+K_{0}\right)^{-1},\\
\mathcal{W_{\textrm{higher}}} & = & \mathcal{C}.\end{array}\label{eq:3.1.16}\end{equation}
 In practice the denominator of the dipole weight $\mathcal{W}_{\textrm{dipole}}$
is modified if we use the modified dipole without the mass terms.
We denote a cut-off on the energy of the photons in the rest frame
of the decay products as $\Omega_{B}$. We will consider the contribution
from the exact higher-order corrections in more detail in the next
section. 
\item There is one remaining complication. The photons are generated in
the frame where the decaying particle enters with momentum equal to
the total photon momentum. However, we wish to apply the energy cut-off
in either the rest frame of the decaying particles, or even the laboratory
frame. There are a number of methods which we could use to achieve
this. The simplest would be to evaluate the YFS form-factor in the
rest frame of the decaying particle or the laboratory frame and veto
any events in which any of the photons are below the cut-off in the
relevant frame. However, this procedure can be inefficient if the
veto removes a large number of events.

We instead choose to use the same procedure as \cite{Jadach:1999vf}.
In this approach we neglect any photons which are below the energy
cut-off in the relevant frame and apply an additional weight {\small \begin{equation}
\mathcal{W}_{\textrm{remove}}=\exp\left(-Y_{12}\left(q_{1},q_{2},\Omega_{B}\right)+Y_{12}\left(q_{1},q_{2},\Omega\right)+Y_{12}\left(p_{1},p_{2},\Omega\right)-Y_{12}\left(p_{1},p_{2},\Omega_{B}\right)\right),\label{eq:3.1.17}\end{equation}
} where $\Omega$ denotes the cut-off on the photon energies in either
the rest frame of the decaying particle or the laboratory frame.

For consistency in defining the infrared region, in applying this
weight we do not apply dipole weights for those photons whose energy
is below the infrared cut-off.%
\footnote{In practice if we neglect the mass terms when generating the crude
distribution we also need to include the weight from (\ref{eq:8.1.8})
for the removed photons as part of the dipole weight in order to take
this into account.%
}

\end{enumerate}

\subsection{Initial-Final Dipole\label{sub:Initial-Final-Dipoles}}

In this subsection we describe our algorithm for dressing decays,
in which a charged particle decays to another charged particle and
a neutral particle. As in the final-final dipole case, the inputs
to the algorithm are the momenta and quantum numbers of the parent
particle and its children, distributed according to the leading-order
differential decay rate.

The situation here is less complicated than for the final-final dipole
case because we can use the neutral decay product to absorb the recoil
due to the photonic radiation. This allows us to simulate the radiation
in the rest frame of the decaying particle. As with the final-final
dipole we must also rescale the three-momentum of the charged particle
in order to have overall energy-momentum conservation.

As in the previous subsection, we begin by manipulating the phase-space
measure in order to factorize off a part of the integrand which can
be interpreted as corresponding to the leading-order process. Taking
$p_{1}$ to be the momentum of the charged particle in the final state
and integrating over the momentum $\left|\vec{p_{1}}\right|$ and
$\vec{p}_{2}$, in the rest frame of the decaying particle gives \begin{equation}
\Gamma=\frac{1}{2M}\sum_{n_{\gamma}=0}^{\infty}\int\mathrm{d}\Phi\textrm{ }\frac{1}{n_{\gamma}!}e^{Y\left(p,p_{1},\Omega\right)}\prod_{i=1}^{n_{\gamma}}\tilde{S}\left(p,p_{1},k_{i}\right)\rho_{\alpha\beta}\mathcal{M}_{\alpha}\mathcal{M}_{\beta}^{*}\mathcal{C}\label{eq:3.2.1}\end{equation}
 where \begin{equation}
\mathrm{d}\Phi=\frac{1}{4\left(2\pi\right)^{2}}\textrm{ }\mathrm{d}\Omega_{p_{1}}\mathrm{d}\Phi_{k}\textrm{ }\frac{\left|\vec{p}_{1}\right|^{2}}{p_{1}^{0}\left(\left|\vec{p}_{1}\right|+\left|\vec{K}\right|\cos\theta_{p_{1}K}\right)+p_{2}^{0}\left|\vec{p}_{1}\right|}.\label{eq:3.2.2}\end{equation}
 This can be rewritten as {\small \begin{equation}
\Gamma=\sum_{n_{\gamma}=0}^{\infty}\int\mathrm{d}\Gamma_{0}\mathrm{d}\Phi_{k}\textrm{ }\frac{\left|\vec{p}_{1}\right|^{2}M}{\left|\vec{q}_{1}\right|p_{1}^{0}\left(\left|\vec{p}_{1}\right|+\left|\vec{K}\right|\cos\theta\right)+\left|\vec{q}_{1}\right|p_{2}^{0}\left|\vec{p}_{1}\right|}\textrm{ }\frac{1}{n_{\gamma}!}e^{Y\left(p,p_{1},\Omega\right)}\prod_{i=1}^{n_{\gamma}}\tilde{S}\left(p,p_{1},k_{i}\right)\mathcal{C}\label{eq:3.2.3}\end{equation}
} where \begin{equation}
\mathrm{d}\Gamma_{0}=\frac{1}{2M}\textrm{ }\mathrm{d}\Omega_{q_{1}}\frac{\left|\vec{q}_{1}\right|}{4\left(2\pi\right)^{2}M}\rho_{\alpha\beta}\mathcal{M}_{\alpha}\mathcal{M}_{\beta}^{*}.\label{eq:3.2.4}\end{equation}
 As before, by not changing the angles of the photons with respect
to the dipole (in this case the charged final-state particle) we have
$\mathrm{d}\Omega_{p_{1}}=\mathrm{d}\Omega_{q_{1}}$. The generation
of the leading-order process $\left(\mathrm{d}\Gamma_{0}\right)$
may proceed prior to, and independently of, the details of QED radiation.
That is to say that no changes need to be made to the existing decay
program, the QED algorithm for initial-final dipoles is universal
in this respect. Momentum conservation and on-shell conditions require
that the momenta, after generation of the photons, are given by\begin{equation}
\begin{array}{lcl}
p & = & q,\\
p_{1} & = & \left(\sqrt{\rho^{2}\vec{q}_{1}^{2}+m_{1}^{2}},\rho\vec{q}_{1}\right),\\
p_{2} & = & \left(M-K_{0}-\sqrt{\rho^{2}\vec{q}^{2}+m_{1}^{2}},-\vec{K}-\rho\vec{q}_{1}\right),\\
K & = & \left(K_{0},\vec{K}\right),\end{array}\label{eq:3.2.5}\end{equation}
 where the rescaling factor $\rho$ is {\footnotesize \begin{equation}
\rho=\frac{-\left|\vec{K}\right|\cos\theta_{1K}\left(\left(p_{1}+p_{2}\right)^{2}+m_{1}^{2}-m_{2}^{2}\right)+\left(M-K_{0}\right)\sqrt{\lambda\left(\left(p_{1}+p_{2}\right)^{2},m_{1}^{2},m_{2}^{2}\right)-4m_{1}^{2}K_{\perp}^{2}}}{2\left|\vec{q}_{1}\right|\left(\left(p_{1}+p_{2}\right)^{2}+K_{\perp}^{2}\right)}\label{eq:3.2.6}\end{equation}
} and $K_{\perp}^{2}=\left|\vec{K}\right|^{2}\sin^{2}\theta_{1K}$.

The crude distribution, is obtained from the exact distribution (\ref{eq:3.2.4})
by dropping the kinematic factor arising from integrating over the
delta function and the higher-order non-soft photon corrections in
$\mathcal{C}$. The momenta in the form factor and dipole functions
are replaced by the values generated from the leading-order decay
$\left(\vec{q}_{1}=-\vec{q}_{2}\right)$ giving the crude distribution
\begin{equation}
\Gamma_{\mathrm{crude}}=\sum_{n_{\gamma}=0}^{\infty}\int\mathrm{d}\Gamma_{0}\mathrm{d}\Phi_{k}\textrm{ }\frac{1}{n_{\gamma}!}e^{Y_{12}\left(q,q_{1},\Omega\right)}\prod_{i=1}^{n_{\gamma}}\tilde{S}\left(q,q_{1},k_{i}\right).\label{eq:3.2.7}\end{equation}
 The dependence of QED part of (\ref{eq:3.2.7}) on the momenta $q$
and $q_{1}$ is overstated here, in the rest frame of the decaying
particle the kinematics are so simple that this part only depends
on the masses $q^{2}$ and $q_{1}^{2}$. Therefore (\ref{eq:3.2.7})
is really a product of two independent integrals. The simplified kinematics
allow the integral over the photon momenta to be performed analytically
giving \begin{equation}
\bar{n}=\int\textrm{ }\frac{\mathrm{d}^{3}k_{i}}{k_{i}^{0}}\textrm{ }\Theta\left(k_{i},\Omega\right)\textrm{ }\tilde{S}\left(q,q_{1},k_{i}\right).\label{eq:3.2.8}\end{equation}
 We therefore obtain \begin{equation}
\Gamma_{\textrm{crude}}=\Gamma_{0}\sum_{n_{\gamma}=0}^{\infty}\frac{1}{n_{\gamma}!}\bar{n}^{n_{\gamma}}e^{-\bar{n}}=\Gamma_{0}.\label{eq:3.2.9}\end{equation}
 In obtaining this we have, as in the final-final dipole case, approximated
$Y\approx-\bar{n}$. Once again we have reduced the width to a simple
Poisson distribution for the photon multiplicity.

The generation of the crude width proceeds in the same way as for
the final-final dipole. First we generate $n_{\gamma}$ according
to the Poisson distribution and then the photon momenta are generated
according to the dipole functions (see Appendix \ref{sub:Final-Final_Generation}).
The form of the rejection weights $\mathcal{W}$ is similar to those
in section \ref{sub:Final-Final-Dipoles} equation (\ref{eq:3.1.15})
with the following changes: \begin{equation}
\begin{array}{lcl}
\mathcal{W_{\textrm{dipole}}} & = & \prod_{i=1}^{n_{\gamma}}\frac{\tilde{S}\left(q,p_{1},k_{i}\right)}{\tilde{S}\left(q,q_{1},k_{i}\right)},\\
\mathcal{W_{\textrm{YFS}}} & = & \frac{e^{Y\left(q,p_{1},\Omega\right)}}{e^{-\bar{n}}},\\
\mathcal{W_{\textrm{Jacobian}}} & = & \frac{\left|\vec{p}_{1}\right|^{2}M}{p_{1}^{0}\left(\left|\vec{p}_{1}\right|+\left|\vec{K}\right|\cos\theta\right)\left|\vec{q}_{1}\right|+p_{2}^{0}\left|\vec{p}_{1}\right|\left|\vec{q}_{1}\right|},\\
\mathcal{W_{\textrm{higher}}} & = & \mathcal{C}.\end{array}\label{eq:3.2.11}\end{equation}
 Unlike the the case of the final-final dipole, we do not need a photon
removal step because the decay is generated in the rest frame of the
decaying particle.

\section{Higher-Order Corrections\label{sec:Higher-Order-Corrections}}

As stated in section \ref{sec:Master-Equation}, the effects of soft
photons (photons with energy below the cut-off $\omega$) have been
included to all orders through the YFS form factor. If one neglects
the infrared residuals in $\mathcal{C}$, the effect of the master
formula and algorithms is, for a given multiplicity, to generate the
QED radiation according to the dipole radiation functions only. This
amounts to approximating matrix elements for the decay $p\rightarrow p_{1}...p_{n}+n_{\gamma}\gamma$
by a product of eikonal factors multiplied by the leading-order matrix
element. Ideally, we wish to include the higher-order effects in $\mathcal{C}$
as far as possible.

Thus far our algorithms only require a set of momenta and their associated
charges. Unfortunately calculating $\mathcal{C}$ exactly to a given
order in $\alpha$ requires knowledge of the matrix elements for the
specific decay process to that order. The structure of \textsf{HERWIG++}
is designed so that if these corrections are known they can be implemented.
However, for the majority of decays these corrections will not be
available and in this case we need to include the remaining enhanced
contributions, \textit{i.e.} the single collinear logarithmic terms.
Depending on the mass scales involved, one can obtain a good approximation
to $\mathcal{C}$ by just including these leading mass singular terms.

In the collinear limit, the squared matrix element for a process including
a massless emission, factorizes into the leading-order squared matrix
element multiplied by a splitting function. The splitting functions
only depend on the spin of the particles involved. Therefore if, in
addition to the momenta and charges, we supply the program with the
spins involved in the decay, we may include the leading non-soft,
collinear logarithms in $\mathcal{C}$ for the real emission contributions.

In addition to affording us a way to include higher-order hard emission
contributions in a universal way, this approach has two further advantages.
Firstly, as we shall describe in more detail in section \ref{sub:Virtual-Corrections},
in this approach we can readily obtain a good approximation to the
virtual corrections. Secondly, given the logarithms associated to
the collinear emissions are universal, they are necessarily gauge
invariant.

\subsection{Real Emission Corrections:$\bar{\beta}_{1}^{1}$\label{sub:Real-Emission-Corrections}}

In the quasi-collinear limit%
\footnote{The quasi-collinear limit is the generalisation of the usual collinear
limit to the case where the emitting particle is massive.%
}, defined in \cite{Catani:2000ef}, the matrix element including the
emission of an additional collinear photon factorizes as \begin{equation}
\left|\mathcal{M}_{R}^{1}\right|^{2}\cong\sum_{i=1}^{n}\frac{e^{2}Z_{i}^{2}}{p_{i}\cdot k}P_{ii}\left|\mathcal{M}\right|^{2},\label{eq:4.1.1}\end{equation}
 where $\mathcal{M}$ is the matrix element for the leading-order
process, $\left|\mathcal{M}_{R}^{1}\right|^{2}$ is the spin-averaged
matrix element with the inclusion of one additional photon, $Z_{i}$
is the charge of the emitting particle, $p_{i}$ is the momentum of
the emitting particle and $k$ is the momentum of the emitted photon.
$P_{ii}$ is the Altarelli-Parisi splitting function for emission
of a photon from particle $i$, its form only depends on the spin
of the emitting particle.

In \cite{Catani:2002hc} these splitting functions were used, together
with the factorization of the matrix element in the soft limit, to
construct so-called \emph{dipole} \emph{splitting functions} for massive
particles. These terms have the correct behaviour in both the soft
and quasi-collinear limits and smoothly interpolate between the two,
\emph{i.e}. they reproduce the massive splitting functions for (quasi-)collinear
emissions and the soft photon, dipole, radiation functions for soft
emissions. We choose to use dipole-like terms based on the expressions
in \cite{Catani:2002hc}, omitting some sub-leading terms which were
included in \cite{Catani:2002hc} to allowed the functions to be analytically
integrated over the phase space of the emitted photon. With the dipole
subtraction terms we may write an approximation for the real emission
matrix element \begin{equation}
\left|\mathcal{M}_{R}^{1}\right|^{2}\approx-e^{2}\sum_{i<j}^{n}Z_{i}\theta_{i}Z_{j}\theta_{j}\left(\mathcal{D}_{ij}+\mathcal{D}_{ji}\right)\left|\mathcal{M}_{n}\right|^{2},\label{eq:4.1.2}\end{equation}
 where indices $i$ and $j$ refer to the two particles forming the
electric dipole and we have applied the conservation of charge $\sum_{i=0}^{n}\theta_{i}Z_{i}=0$.

We adopt the convention that the first index on $\mathcal{D}_{ij}$
refers to the particle in the dipole which is considered to be emitting
the photon, while the second index refers to the so-called \emph{spectator}
particle. From here we may write down a leading collinear approximation
for the infrared finite residual $\bar{\beta}_{1}^{1}$\begin{equation}
\bar{\beta}_{1}^{1}=-\frac{\alpha}{4\pi^{2}}\textrm{ }\rho_{\alpha\beta}\mathcal{M}_{\alpha}\mathcal{M}_{\beta}^{*}\textrm{ }\sum_{i<j}^{n}Z_{i}\theta_{i}Z_{j}\theta_{j}\left[\bar{\mathcal{D}}_{ij}+\bar{\mathcal{D}}_{ji}]\right],\label{eq:4.1.3}\end{equation}
 where the $\bar{\mathcal{D}}_{ij}$ are the infrared subtracted counterparts
of $\mathcal{D}_{ij}$, \begin{equation}
\bar{\mathcal{D}}_{ij}=\mathcal{D}_{ij}-\frac{1}{p_{i}\cdot k}\left[\frac{2p_{i}\cdot p_{j}}{(p_{i}+p_{j})\cdot k}-\frac{m_{i}^{2}}{(p_{i}\cdot k)}\right].\label{eq:4.1.4}\end{equation}

For the case that both the emitter and spectator are in the final
state, the dipole terms are given by%
\footnote{It is not possible to construct a quasi-collinear limit for the spin-1
splitting function, for a massive vector particle, in which the massless
limit can be taken. We therefore use the corresponding expression
for the massless dipole splitting function, augmented by a mass term
which produces the correct behaviour in the soft limit.%
}\begin{equation}
\begin{array}{lcll}
\mathcal{D}_{ij} & = & \frac{1}{p_{i}.k}\left[\frac{2p_{i}.p_{j}}{\left(p_{i}+p_{j}\right).k}-\frac{m_{i}^{2}}{p_{i}.k}\right] & \textrm{spin 0},\\
 & = & \frac{1}{p_{i}.k}\left[\frac{p_{j}.k}{\left(p_{i}+k\right).p_{j}}+\frac{2p_{i}.p_{j}}{\left(p_{i}+p_{j}\right).k}-\frac{m_{i}^{2}}{p_{i}.k}\right] & \textrm{spin }\frac{1}{2},\\
 & = & \frac{1}{p_{i}.k}\left[\frac{2\left(p_{j}.k\right)\left(p_{i}.p_{j}\right)}{\left(\left(p_{i}+k\right).p_{j}\right)^{2}}+\frac{2p_{j}.k}{\left(p_{j}+k\right).p_{i}}+\frac{2p_{i}.p_{j}}{\left(p_{i}+p_{j}\right).k}-\frac{m_{i}^{2}}{p_{i}.k}\right] & \textrm{spin }1.\end{array}\label{eq:4.1.5}\end{equation}
 For the case that the dipole is comprised of the decaying particle
(which we shall denote by index $j$) and one of its children, the
$\mathcal{D}_{ij}$ functions for emissions from the children are
taken to be the same as in (\ref{eq:4.1.5}). However, for the decaying
particle, we assume that it is sufficiently massive for us to neglect
collinear enhancements, giving the following dipole function\begin{equation}
\begin{array}{lcll}
\mathcal{D}_{ji} & = & \frac{1}{p_{j}.k}\left[\frac{2p_{i}.p_{j}}{\left(p_{i}+p_{j}\right).k}-\frac{m_{j}^{2}}{p_{j}.k}\right] & \textrm{spin }0,\frac{1}{2},1\end{array}.\label{eq:4.1.6}\end{equation}
 Since the effects of collinear radiation from the decaying particle
are neglected, only soft emissions are taken into account and so this
dipole function is independent of the particle's spin. Furthermore,
the infrared subtracted dipole splitting function, with the parent
particle as the emitter, is identically zero: $\bar{\mathcal{D}}_{ji}\equiv0$.

In the soft limit these expressions reproduce the expected eikonal
result\begin{equation}
\mathcal{D}_{ij}+\mathcal{D}_{ji}=-\left(\frac{p_{i}}{p_{i}.k}-\frac{p_{j}}{p_{j}.k}\right)^{2}\label{eq:4.1.7}\end{equation}
 and in the collinear limit the $\mathcal{D}_{ij}$ equals the quasi-collinear
splitting functions given in \cite{Catani:2000ef}.

\subsection{Virtual Corrections: $\bar{\beta}_{0}^{1}$\label{sub:Virtual-Corrections}}

At present we have only implemented virtual corrections for two special
cases in the \textsf{SOPHTY} code, those of initial-final and final-final
dipoles with (relativistic) fermions in the final state, as in $\mathrm{W}$
and $\mathrm{Z}$ boson decays. These corrections turn out to have
a negligible effect on \emph{distributions}, compared to those of
the real corrections. This is seen to be the case even for $\mathrm{W}\rightarrow\mathrm{e}^{+}\nu_{\mathrm{e}}$
and $\mathrm{Z}\rightarrow\mathrm{e}^{+}\mathrm{e}^{-}$ decays where
one expects such effects to be greatest.

As with the real emission we try to work in a universal way, without
referring to the details of the matrix elements, using the leading
log approximation.

For both the case of the final-final dipole and the initial-final
dipole the relevant virtual processes are represented by the lowest-order
diagram with a photon joining the dipole constituents. On dimensional
grounds, the large, leading logarithms of QED will be logarithms of
$M^{2}/m_{i}^{2}$. Also, if we regularize the infrared divergences
by introducing a fictitious photon mass $m_{\gamma}$, logarithms
of $M^{2}/m_{\gamma}^{2}$ and $m_{i}^{2}/m_{\gamma}^{2}$ are possible.

The infrared divergences from virtual corrections, must cancel the
infrared divergences arising from the soft region of the photon phase
space in the real emission process \cite{Bloch:1937pw}. Likewise,
terms diverging as $m_{i}^{2}\rightarrow0$, so called \emph{mass/collinear
divergences}, must also cancel between the real emission corrections
and their virtual counterparts, this is the KLN theorem \cite{Kinoshita:1962ur,Lee:1964is}.
Using the fact that the $m_{\gamma}\rightarrow0$ and $m_{i}^{2}\rightarrow0$
divergent logarithms have to cancel in this way, we can construct
the leading log approximation to the loop integrals.

To obtain the leading soft and collinear contributions to the virtual
terms we therefore return to the soft and collinear approximation
that was used for the real emission matrix element (\ref{eq:4.1.2}).
We calculate the leading logarithms arising in the real emission contribution
by integrating the full dipole function over the \textit{full} phase
space for the emission of the photon \textit{i.e.} both the soft $k_{0}<\omega$
and hard $k_{0}\geq\omega$ regions as was done in \cite{Catani:2002hc}.
Performing the relevant integrals and taking the small mass limit
gives the contribution of the virtual terms for the different types
of dipole \begin{equation}
\begin{array}{lcll}
\left.\mathrm{d}\Gamma\right|_{\mathrm{LL}} & = & \frac{\alpha}{\pi}\left(2\left(\ln\left(\frac{M}{m_{i}}\right)-1\right)\ln\left(\frac{m_{\gamma}}{M}\right)+\ln^{2}\left(\frac{M}{m_{i}}\right)+\frac{1}{2}\ln\left(\frac{M}{m_{i}}\right)\right)\mathrm{d}\Gamma_{0} & \textnormal{(initial-final),}\\
\left.\mathrm{d}\Gamma\right|_{\mathrm{LL}} & = & \frac{\alpha}{\pi}\left(2\left(\ln\left(\frac{M^{2}}{m_{i}^{2}}\right)-1\right)\ln\left(\frac{m_{\gamma}}{M}\right)+\frac{1}{2}\ln^{2}\left(\frac{M^{2}}{m_{i}^{2}}\right)+\frac{1}{2}\ln\left(\frac{M^{2}}{m_{i}^{2}}\right)\right)\mathrm{d}\Gamma_{0} & \textnormal{(final-final)}.\end{array}\label{eq:4.2.5}\end{equation}
These expressions agree, at the level of large logarithms, with those
obtained by direct calculation in \cite{Marciano:1974vg} and \cite{Bonneau-Martin}.
From here we see that the $\bar{\beta}_{0}^{1}$ functions we require
are, for the initial-final dipole\begin{equation}
\bar{\beta}_{0}^{1}=\frac{\alpha}{2\pi}\ln\left(\frac{M^{2}}{m_{i}^{2}}\right)\bar{\beta}_{0}^{0},\label{eq:4.2.6}\end{equation}
 and for the final-final dipole,\begin{equation}
\bar{\beta}_{0}^{1}=\frac{\alpha}{\pi}\ln\left(\frac{M^{2}}{m_{i}^{2}}\right)\bar{\beta}_{0}^{0}.\label{eq:4.2.7}\end{equation}
 For resonant $\mathrm{Z}\rightarrow\mathrm{e}^{+}\mathrm{e}^{-}$
processes this number is around 6\%, dropping to around 3\% for resonant
$\mathrm{Z}\rightarrow\mu^{+}\mu^{-}$ processes. The extension to
other cases is obvious, it simply requires the use of the scalar and
vector splitting functions instead.

\section{Results\label{sec:Results}}

In this section we discuss the results from the \textsf{SOPHTY} program
as implemented in \textsf{HERWIG++}. In order to test the algorithm
we will consider both leptonic $\mathrm{Z}$ and $\mathrm{W}$ boson
decays, due to their phenomenological importance. In addition we will
consider a number of important meson decays to demonstrate the application
of the program to hadronic decays. We reiterate that our approach
simulates the soft photon corrections in the leading log approximation
which depends on nothing more than the momenta and charges of the
primary decay products, and simulation of hard collinear photons merely
requires additional spin information. This being the case these examples
represent a general test of our methods.

\begin{figure}[t]
\includegraphics[%
  clip,
  width=0.34\textwidth,
  angle=90]{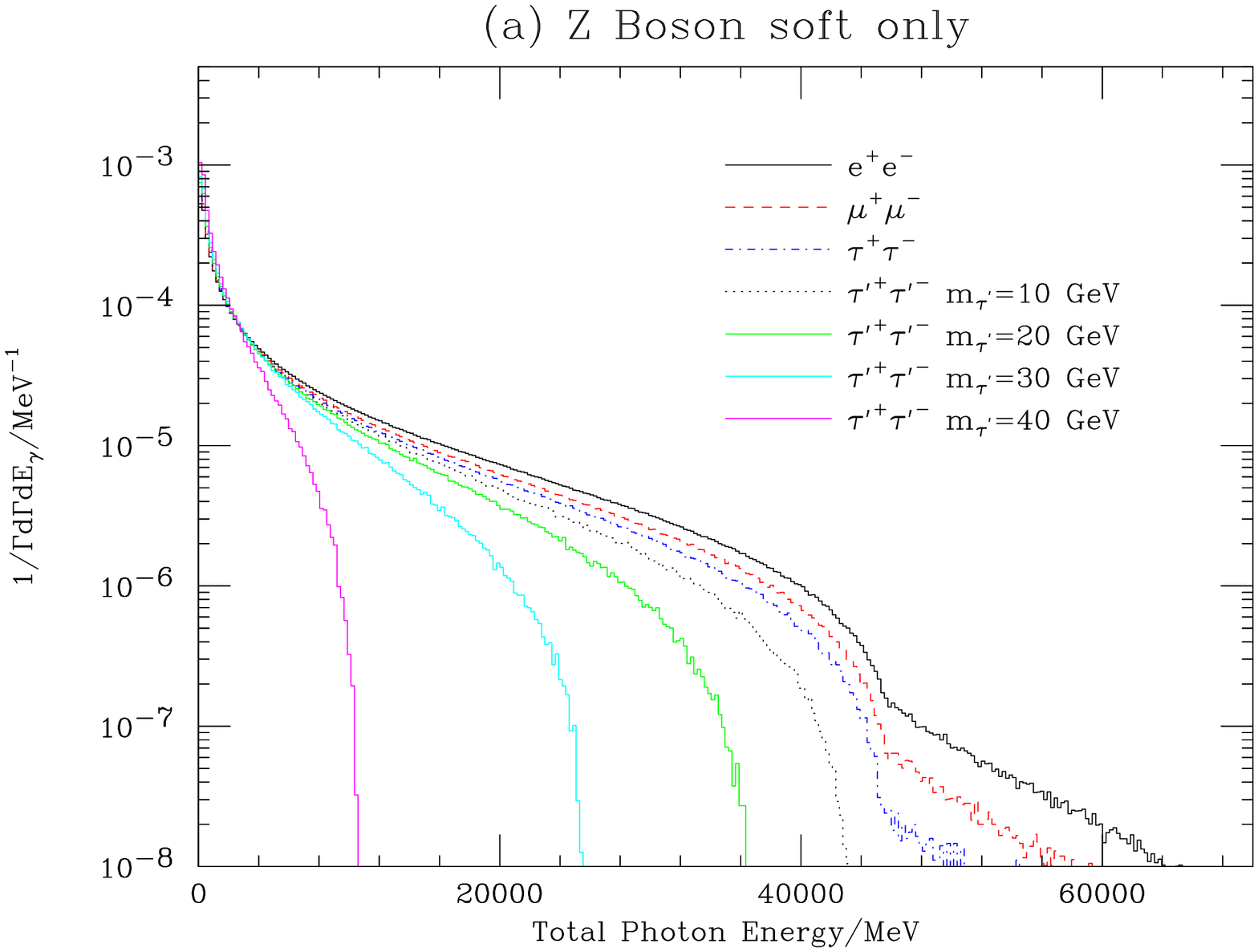}\hfill{}\includegraphics[%
  clip,
  width=0.34\textwidth,
  angle=90]{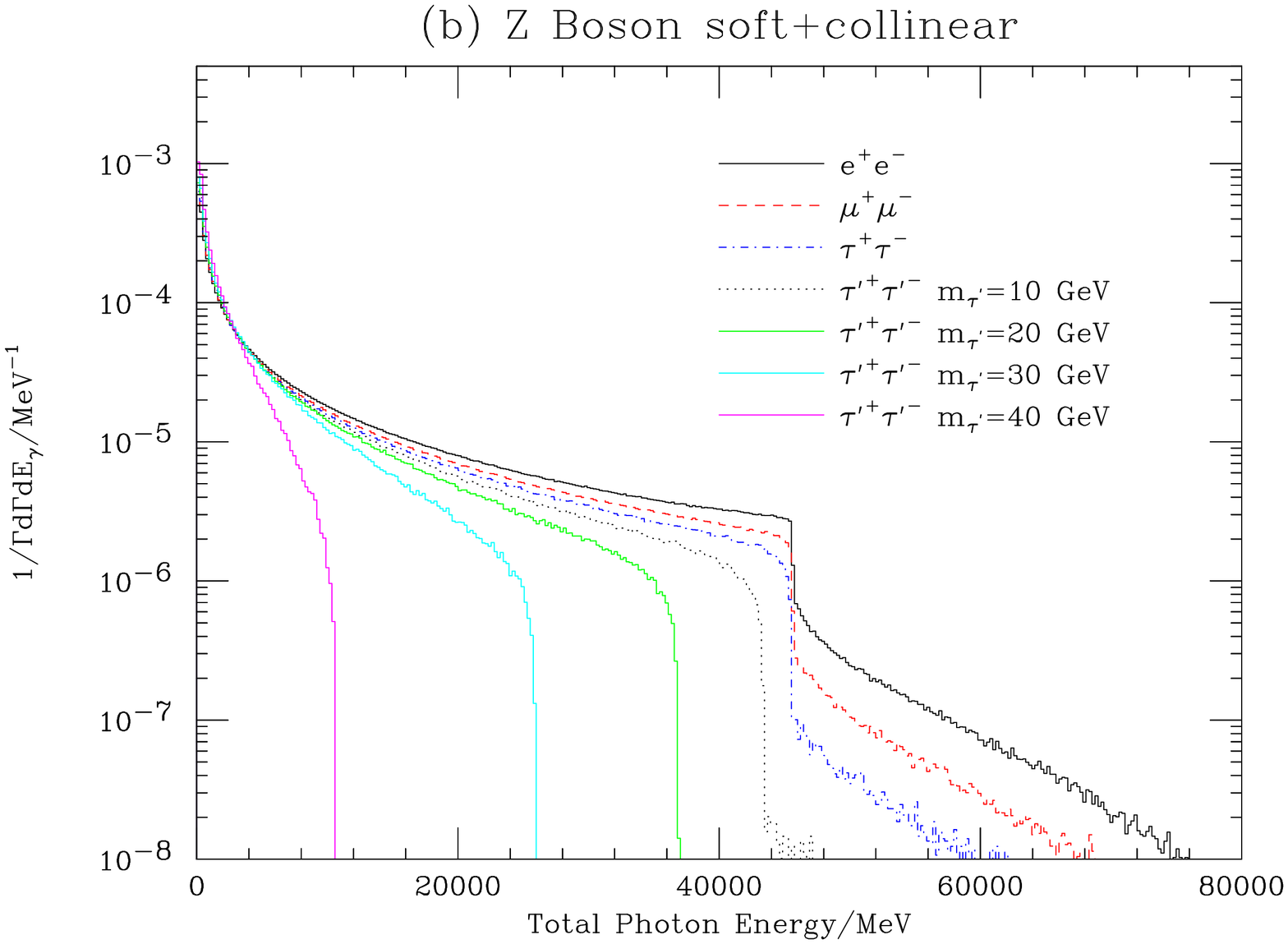}

\caption{\label{cap:Z_tot_ph_energy}The total energy $\left(K_{0}\right)$
of the photons radiated in $\mathrm{Z}$ boson decays to leptons:
(a) shows the $K_{0}$ spectrum for the case that no infrared residuals
are considered $\left(\mathcal{C}=1\right)$; (b) shows the effect
of including the collinear approximation for the $\mathcal{O}\left(\alpha\right)$
residual $\bar{\beta}_{1}^{1}$.}
\end{figure}

\begin{figure}[!h]
\includegraphics[%
  width=0.34\textwidth,
  angle=90]{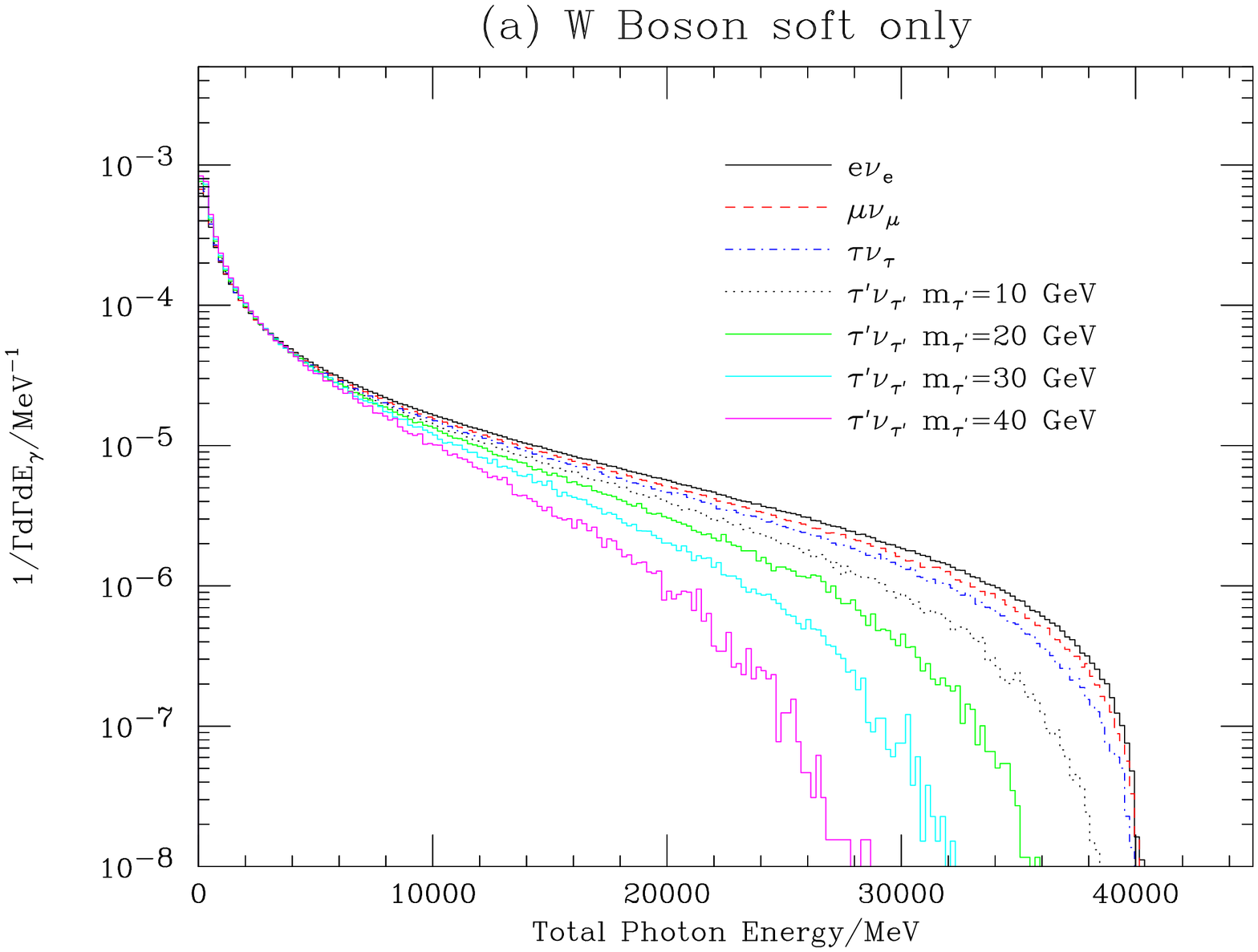}\hfill{}\includegraphics[%
  width=0.34\textwidth,
  angle=90]{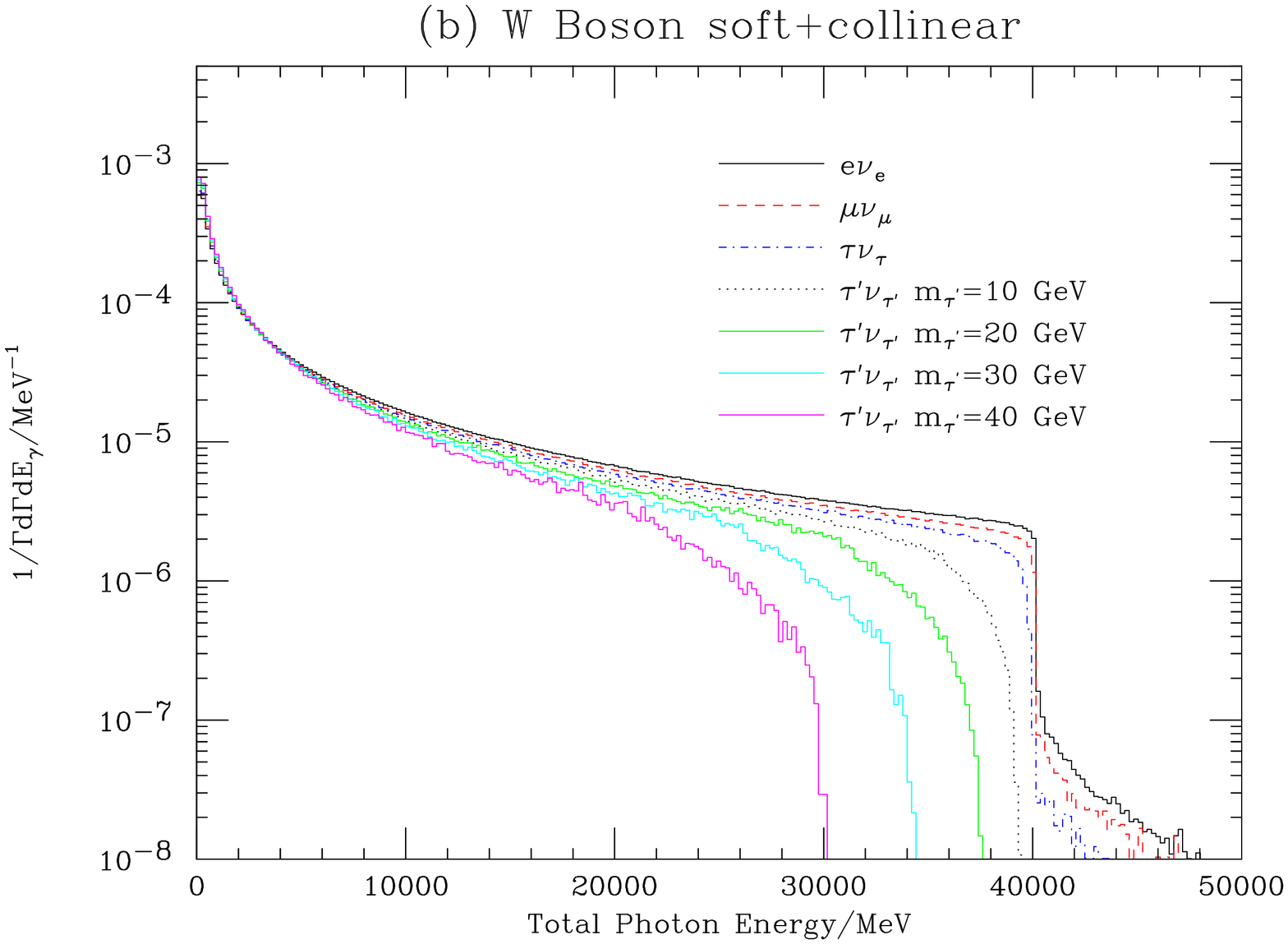}

\caption{\label{cap:W_tot_ph_energy}The total energy $\left(K_{0}\right)$
of the photons radiated in $\mathrm{W}$ boson decays to leptons;
(a) shows the $K_{0}$ spectrum for the case that no infrared residuals
are considered $\left(\mathcal{C}=1\right)$; (b) shows the effect
of including the collinear approximation for the $\mathcal{O}\left(\alpha\right)$
residual $\bar{\beta}_{1}^{1}$.}
\end{figure}

The key distribution produced by the simulation is the total photon
energy spectrum $\left(K_{0}\right)$. This is shown for $\mathrm{Z}$
decays in figure \ref{cap:Z_tot_ph_energy} and for $\mathrm{W}$
decays in figure \ref{cap:W_tot_ph_energy}. We have considered a
large range of masses for the decay products, including a fictitious
heavy lepton $\left(\tau^{\prime}\right)$, to check for numerical
instabilities and other irregular behaviour. For each type of decay
we show the results of our algorithm including only soft photon effects
and also with the dipole approximation for hard radiation. In all
cases the amount of radiation is seen to decrease smoothly as the
mass of the decay products increases, this can be understood from
simple phase-space considerations. 

\begin{figure}[t]
\begin{center}\includegraphics[%
  width=0.34\textwidth,
  angle=90]{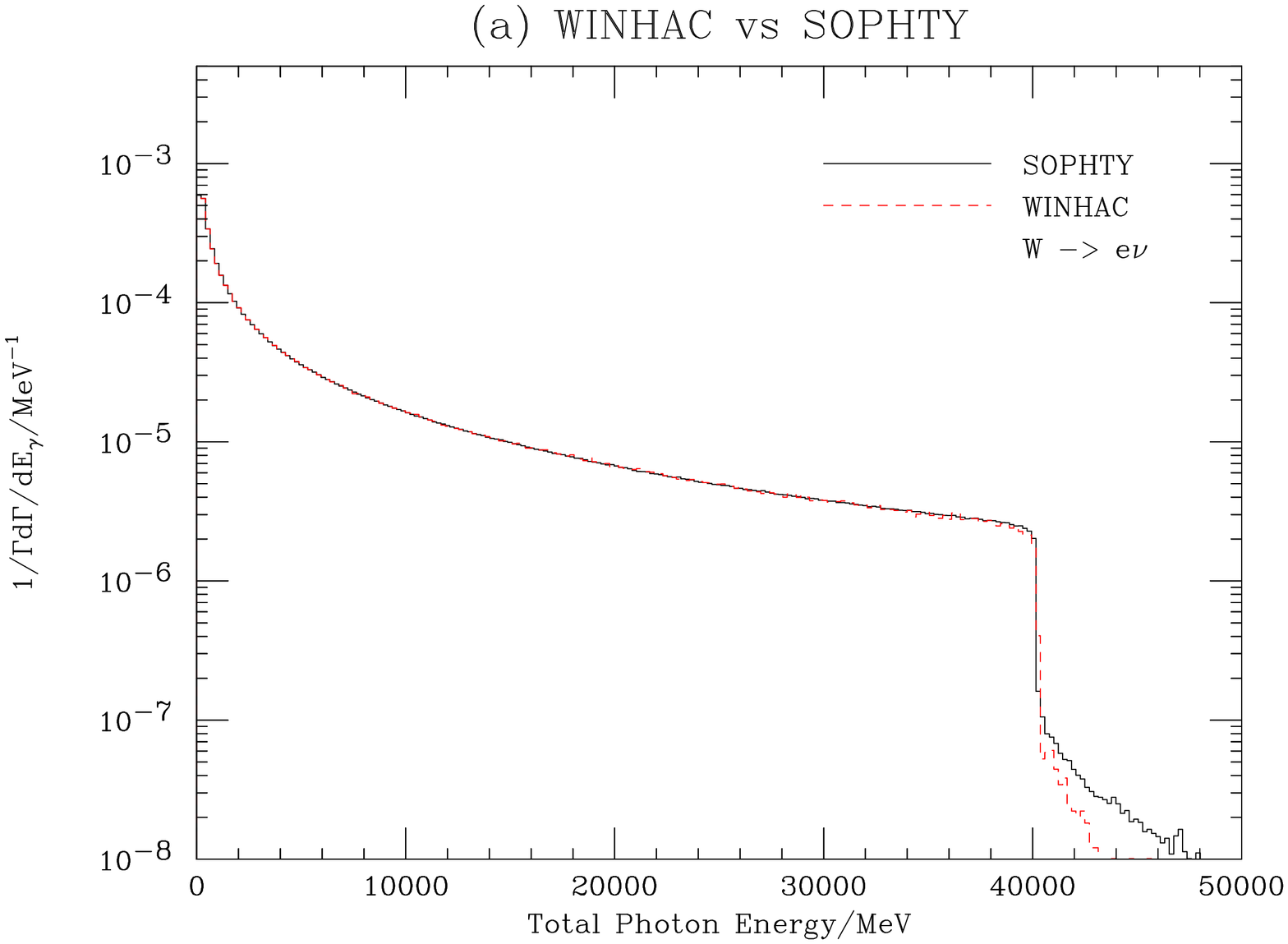}\hfill{}\includegraphics[%
  width=0.34\textwidth,
  angle=90]{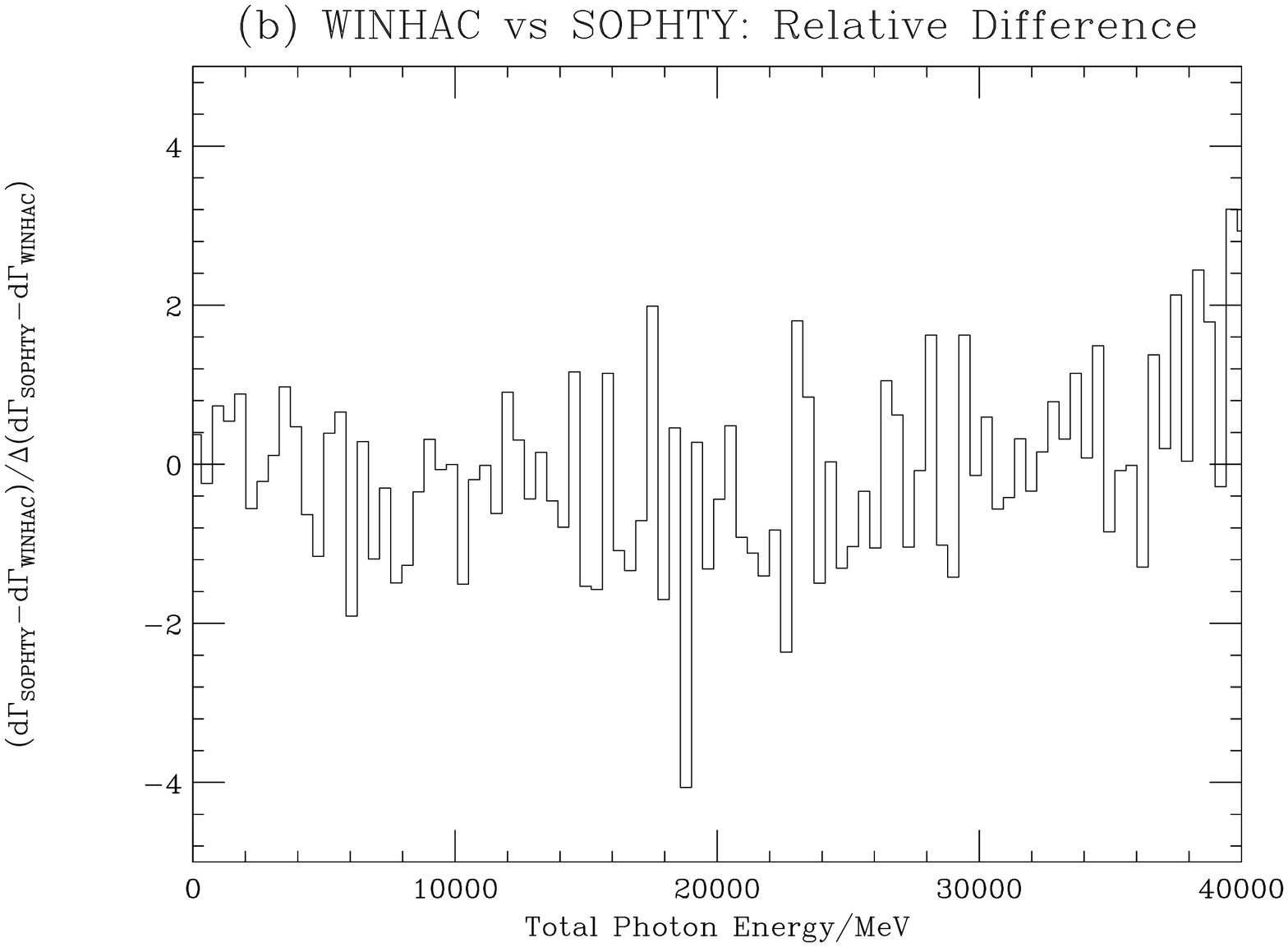}\end{center}

\caption{\label{cap:wh_v_sophty_2}The total energy $\left(K_{0}\right)$
of the photons radiated in $\mathrm{W}^{\pm}\rightarrow\mathrm{e}^{\pm}\nu_{e}/\bar{\nu_{e}}$
decays. In figure (a) the red histogram was generated using the \textsf{WINHAC}
\cite{Placzek:2003zg} simulation, including exact $\mathcal{O}\left(\alpha\right)$
real emission corrections to the $\mathrm{W}^{\pm}$ decay, while
the black line was generated using the \textsf{SOPHTY} module for
QED radiation in \textsf{HERWIG++}. In figure (b) we show the
difference between the spectra shown in (a) divided by the associated
statistical error. The discrepancy in the
region beyond 40 GeV is exclusively comprised of events with at least
two non-soft photons, which neither program is designed to model well.}
\end{figure}

One can also see that the inclusion of the dipole splitting functions
$\left(\mathcal{D}_{ij}\right)$ leads to an enhancement of hard photon
radiation. This enhancement is most prominent for lighter decay products,
for the heavier decay products the effect is negligible. Again, this
is to be expected as the mass of the emitting particle is known to
screen the collinear divergence, this can be seen by considering,
for instance, the massive splitting functions in \cite{Catani:2000ef}.

Including the hard collinear enhancements also reveals a kink in the
total photon energy spectrum. This kink occurs at a kinematic endpoint,
beyond it all events must contain at least two photons which recoil
against each other, hence the histograms drop beyond this value. Looking
in this two photon region we also see that the photon multiplicity
increases as the mass of the primary decay products decreases.

Changing the spin of the primary decay products does not affect the
soft distributions in figures \ref{cap:Z_tot_ph_energy}a and \ref{cap:W_tot_ph_energy}a,
it does however, influence the other distributions where hard collinear
photon effects are introduced. The program uses the other splitting
functions in (\ref{eq:4.1.5}) to account for this, although in the
case that the decaying particle is a scalar there will be no collinear
enhancement since in this case $\bar{\mathcal{D}}_{ij}\equiv0$.

In figure \ref{cap:wh_v_sophty_2} we compare the total photon energy
spectrum for $\mathrm{W}\rightarrow\mathrm{e}\nu_{\mathrm{e}}$ decays
as generated by our program and that of the \textsf{WINHAC} generator.
The agreement is seen to be quite good except in the region beyond
the kink at around 40 GeV. As mentioned earlier, this region is populated
exclusively by events with at least two hard photons. Consequently
neither simulation expects to model this accurately. A correct modelling
of this region will require the calculation of the infrared residuals
$\left(\mathcal{C}\right)$ to $\mathcal{O}\left(\alpha^{2}\right)$.
This extension may be implemented in future versions of the program.
We note that \textsf{WINHAC} was been independently compared with
another simulation of the charged Drell-Yan process, \textsf{HORACE,}
in \cite{CarloniCalame:2004qw}, where good numerical agreement between
the different approaches to QED radiation was recorded.

\begin{figure}[t]
\begin{center}\includegraphics[%
  width=0.34\textwidth,
  angle=90]{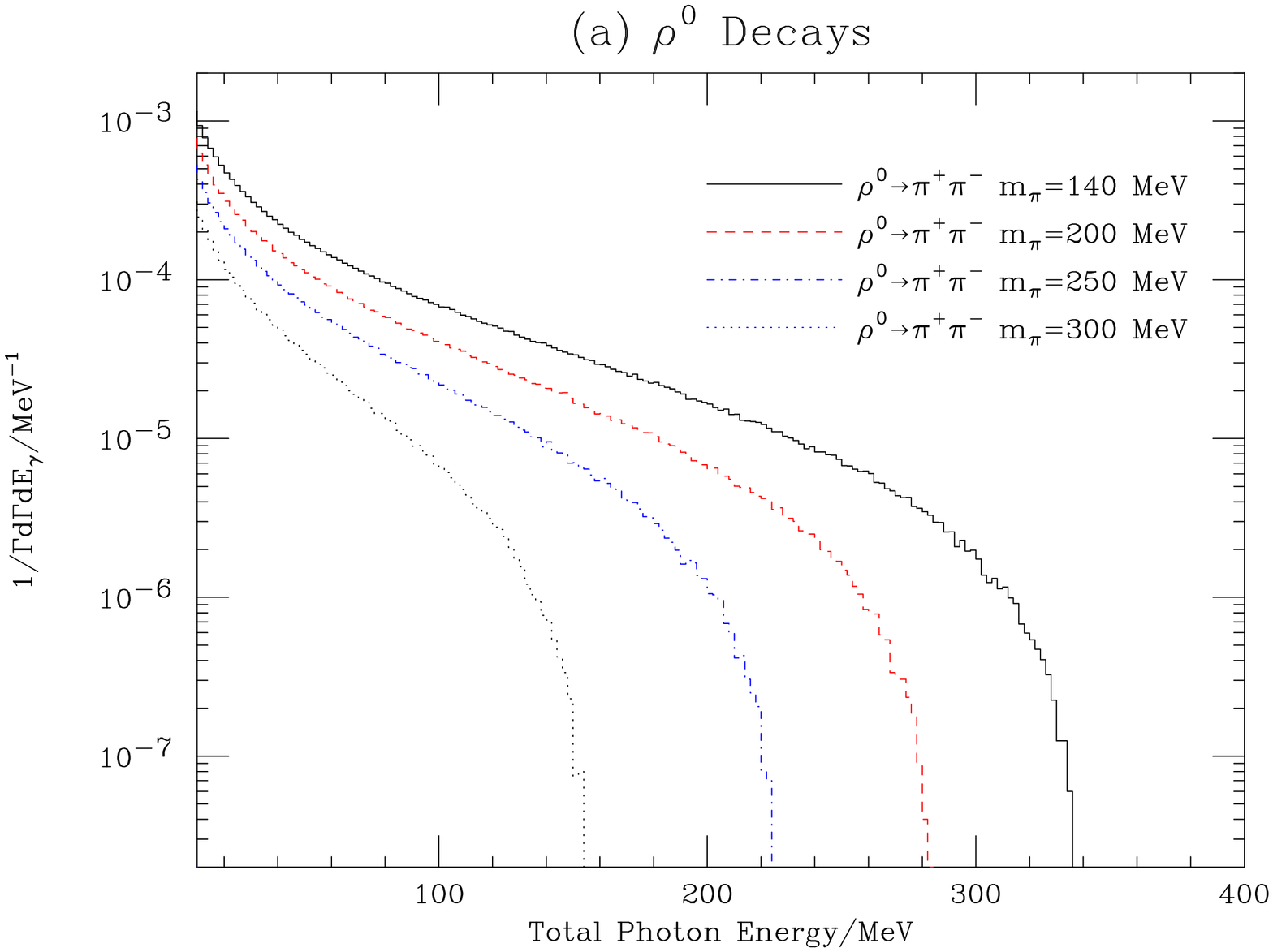}\hfill{}\includegraphics[%
  width=0.34\textwidth,
  angle=90]{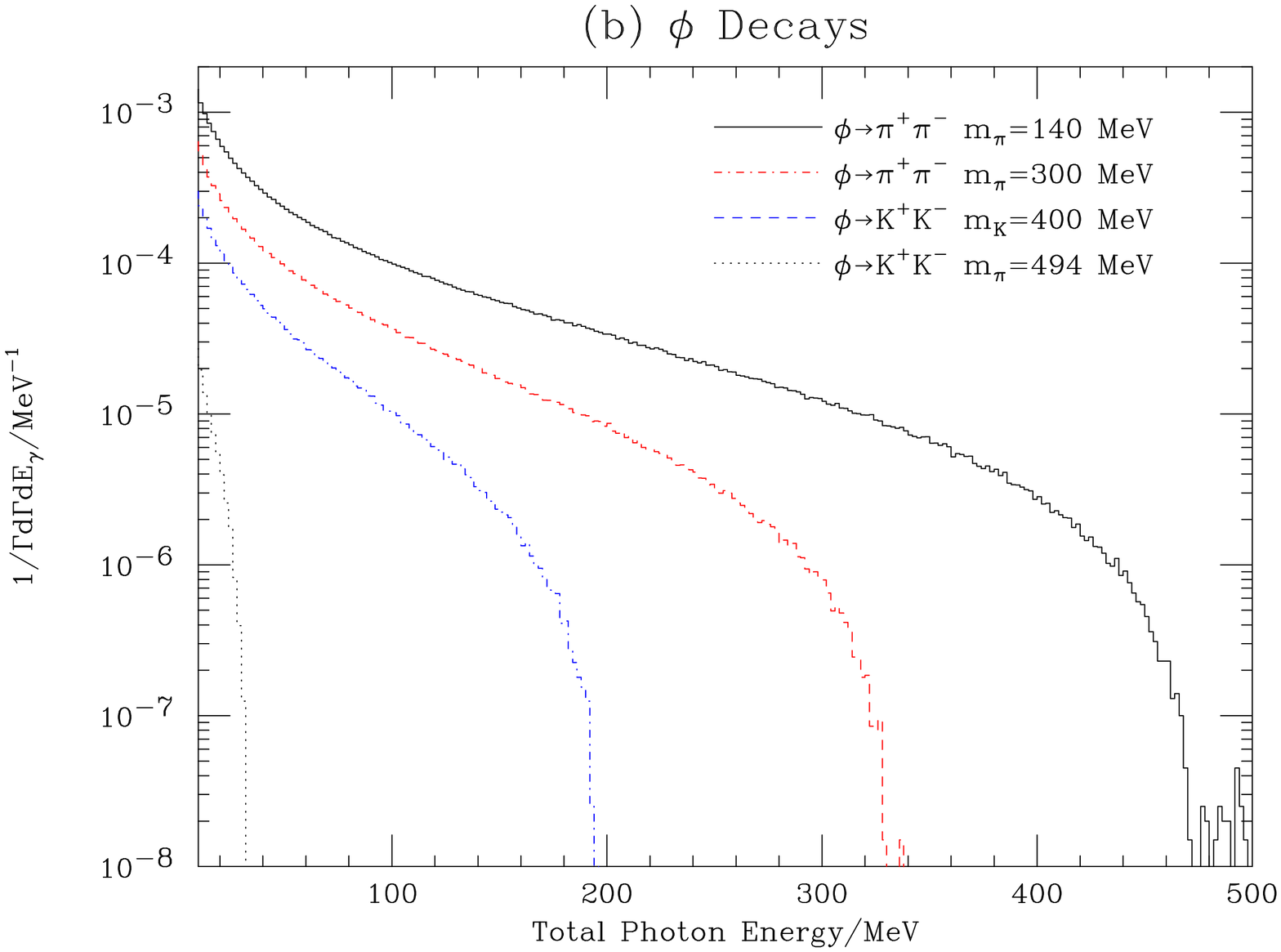}\\
\includegraphics[%
  width=0.34\textwidth,
  angle=90]{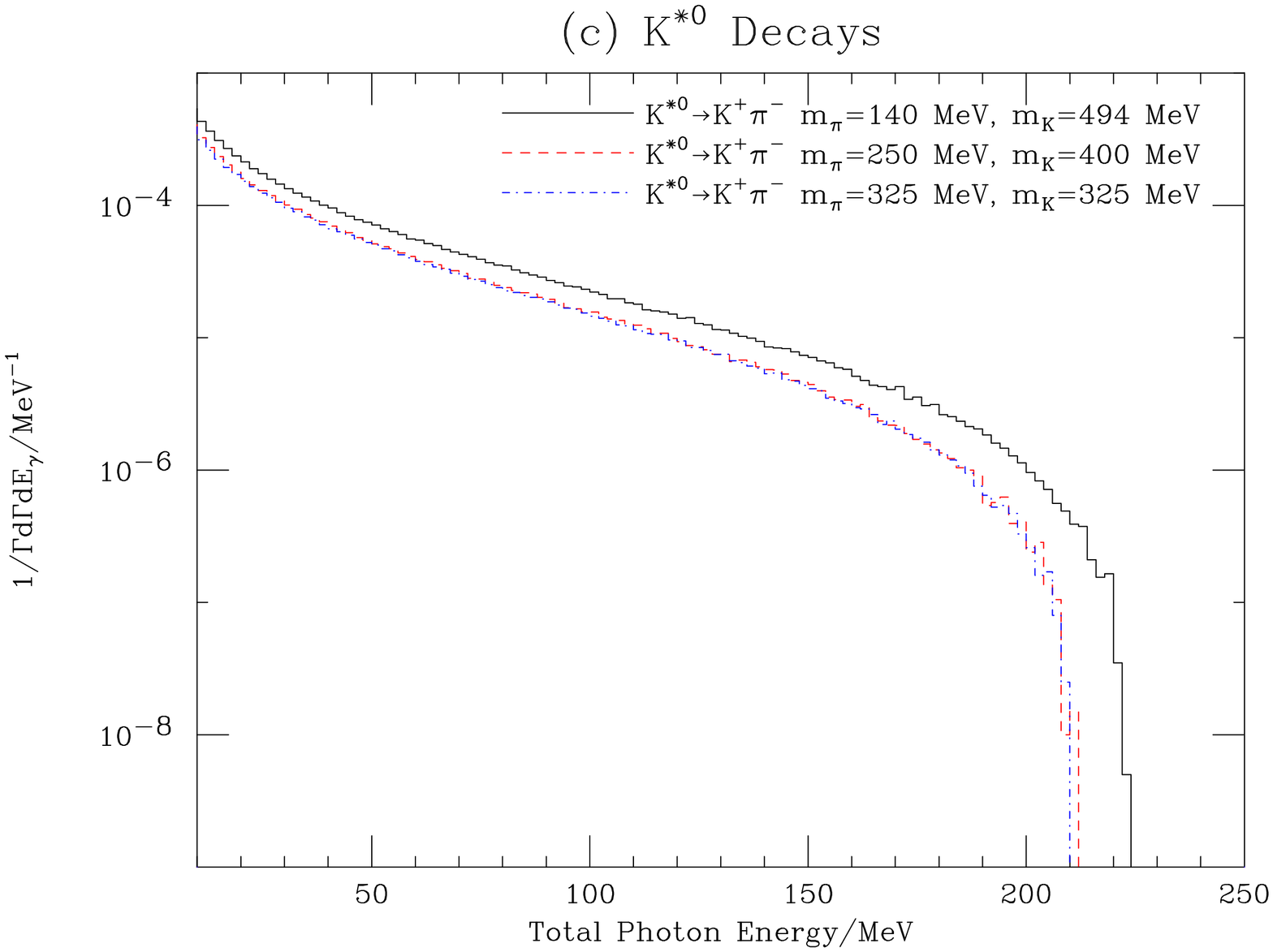}\hfill{}\includegraphics[%
  width=0.34\textwidth,
  angle=90]{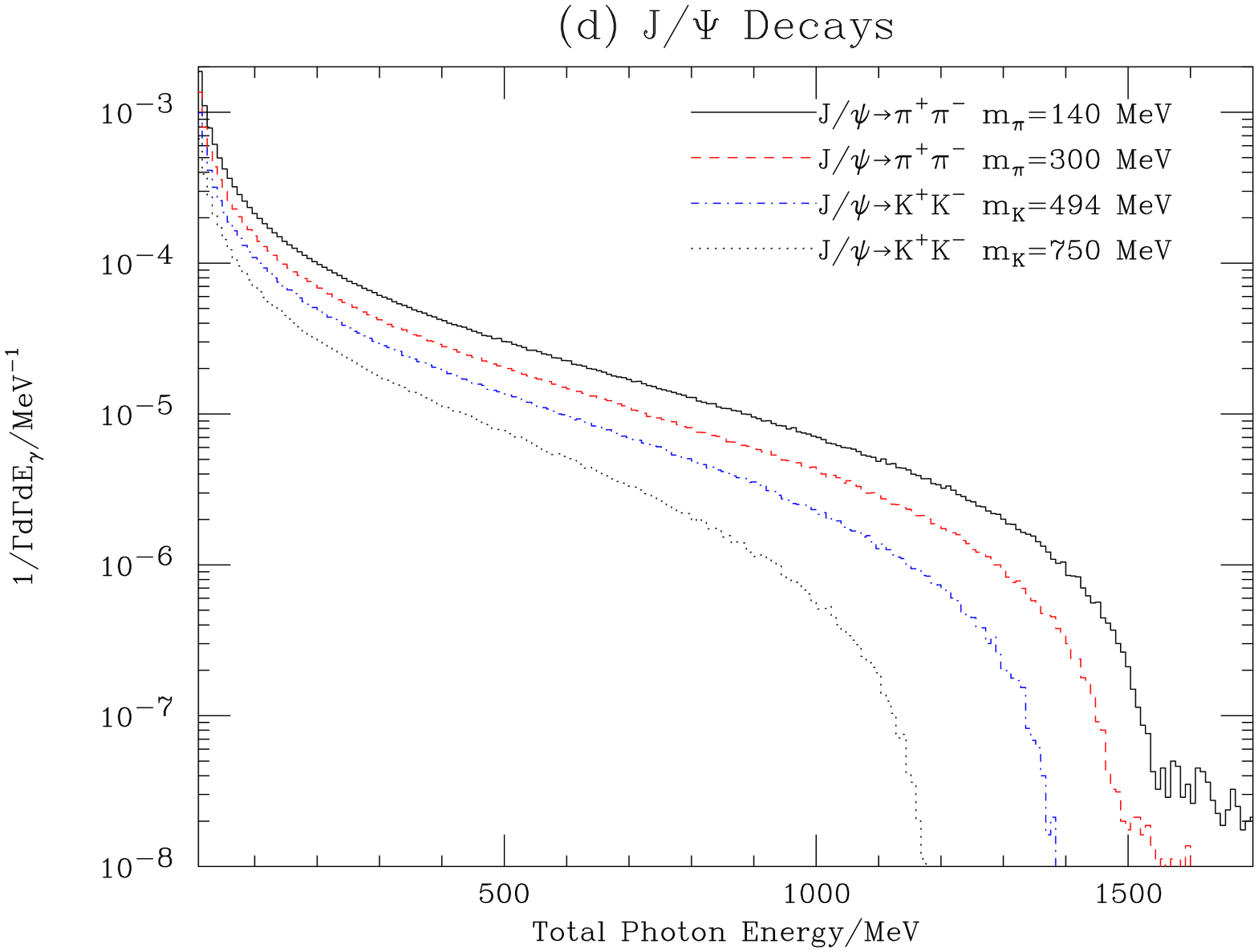}\end{center}

\caption{The total energy $\left(K_{0}\right)$ of the photons radiated in
the decays of neutral vector mesons to pseudoscalar mesons for a number
of different decays: (a) $\rho\rightarrow\pi^{+}\pi^{-}$; (b) $\phi\rightarrow\pi^{+}\pi^{-}$
and $\phi\rightarrow{\textrm{K}}^{+}{\textrm{K}}^{-}$; (c) ${\textrm{K}}^{*0}\rightarrow{\textrm{K}}^{\pm}\pi^{\mp}$;
(d) ${\textrm{J}}/\psi\rightarrow\pi^{+}\pi^{-}$ and ${\textrm{J}}/\psi\rightarrow{\textrm{K}}^{+}{\textrm{K}}^{-}$.
In addition to using the real physical masses of the decay products
we have included the effect of varying the masses of the decay products.}

\label{fig:neutral_decays}
\end{figure}

The total energy of the photons radiated in the decays of some neutral
vector mesons is shown in figure~\ref{fig:neutral_decays}. Here
we see that the energy distribution shows a behaviour that qualitatively
resembles that seen in the case of the $\mathrm{Z}$ boson. In this
case, as the decay products are pseudoscalar mesons, there is no effect
from including the collinear approximation for the radiation. In addition
figure~\ref{fig:neutral_decays}c shows the radiation for an example,
${\textrm{K}}^{*0}\rightarrow{\textrm{K}}^{\pm}\pi^{\mp}$, with unequal
masses for the decay products. Here we see that the lighter decay
product is responsible for the more energetic photons, as we increase
its mass (for illustrative purposes) the distribution tends toward
lower energies.

\begin{figure}[!h]
\begin{center}\includegraphics[%
  width=0.34\textwidth,
  angle=90]{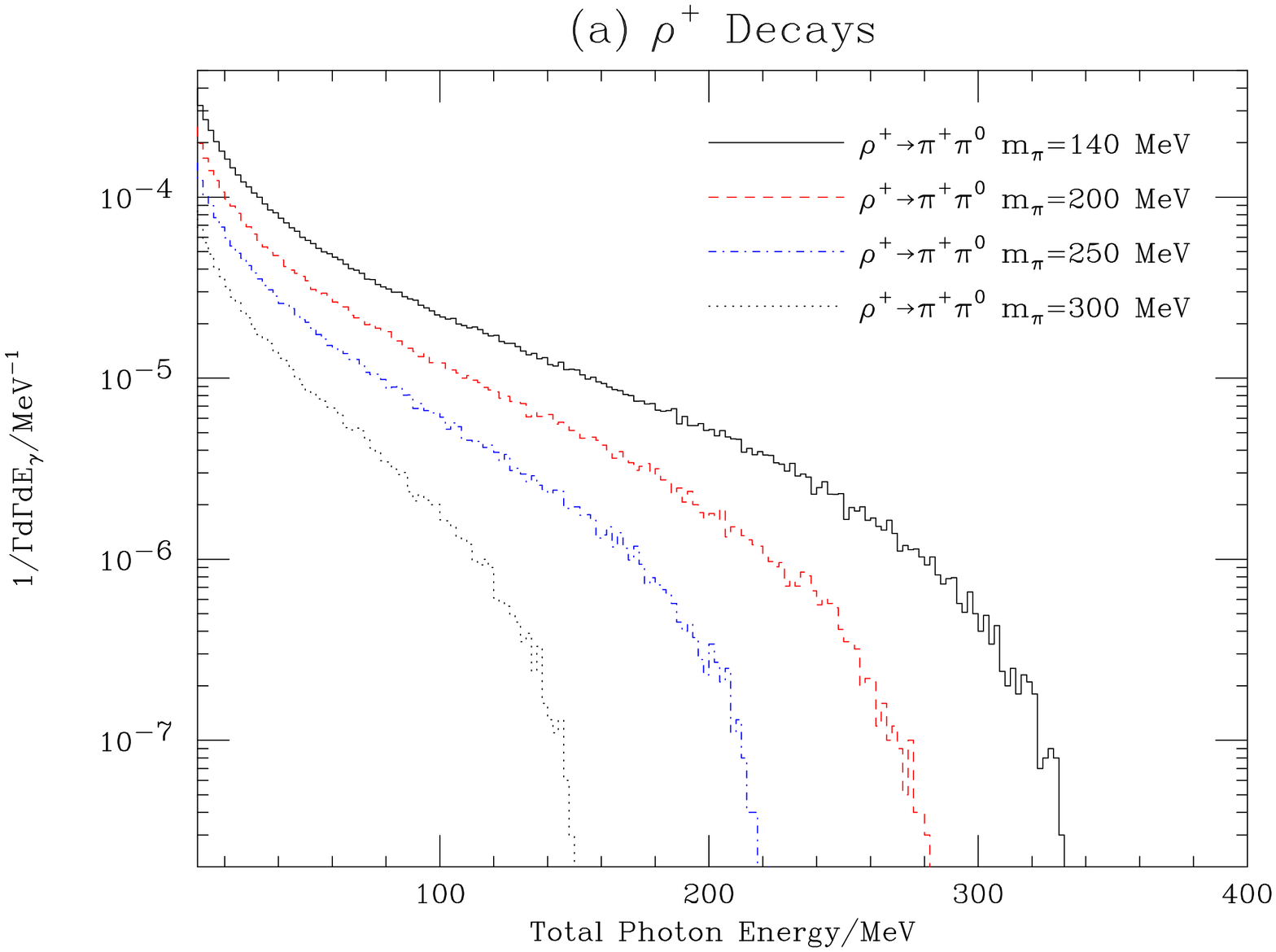}\hfill{}\includegraphics[%
  width=0.34\textwidth,
  angle=90]{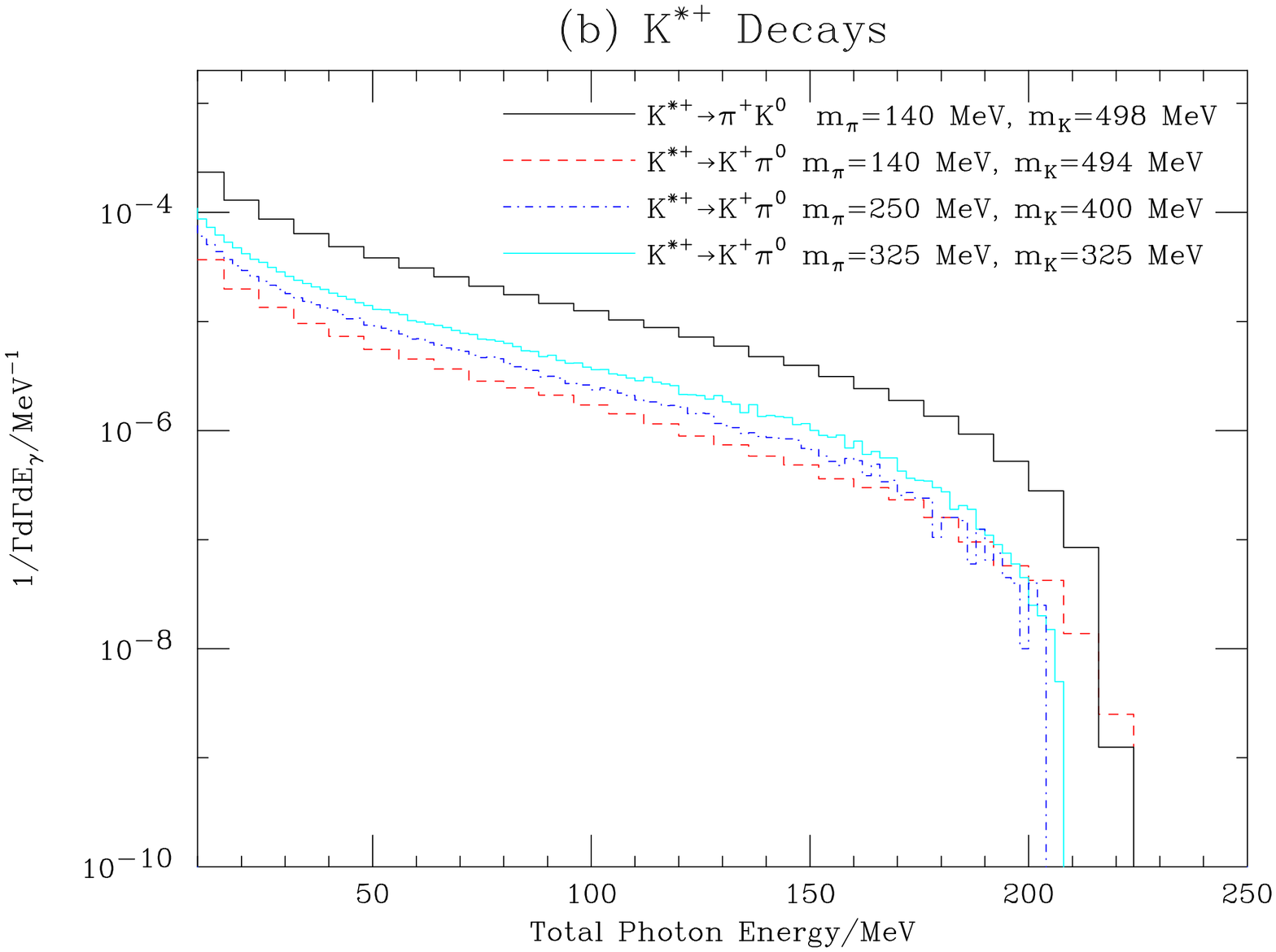}\end{center}

\caption{The total energy $\left(K_{0}\right)$ of the photons radiated in
the decays of charged vector mesons to pseudoscalar mesons for the
decays: (a) $\rho^{\pm}\rightarrow\pi^{\pm}\pi^{0}$; (b) ${\textrm{K}}^{*\pm}\rightarrow{\textrm{K}}^{\pm}\pi^{0}$
and ${\textrm{K}}^{*\pm}\rightarrow{\textrm{K}}^{0}\pi^{\pm}$. In
addition to using the real physical masses of the decay products we
have included the effect of varying the masses of the decay products. }

\label{fig:charged_decays}
\end{figure}

The total energy of the photonic radiation in the decays of some charged
vector mesons is shown in figure~\ref{fig:charged_decays}. As expected,
for these decays the distributions behave in a similar way to those
of the $\mathrm{W}$ boson, since they involve the same type of dipole.
As with the neutral vector meson decays there is no effect from including
the collinear approximation for the photon radiation as the decay
products are pseudoscalar mesons $\left(\bar{\mathcal{D}}_{ij}\equiv0\right)$.
The ${\textrm{K}}^{*\pm}$ decays show the effect of having unequal
mass decay products.

\begin{figure}[!h]
\begin{center}\includegraphics[%
  width=0.34\textwidth,
  angle=90]{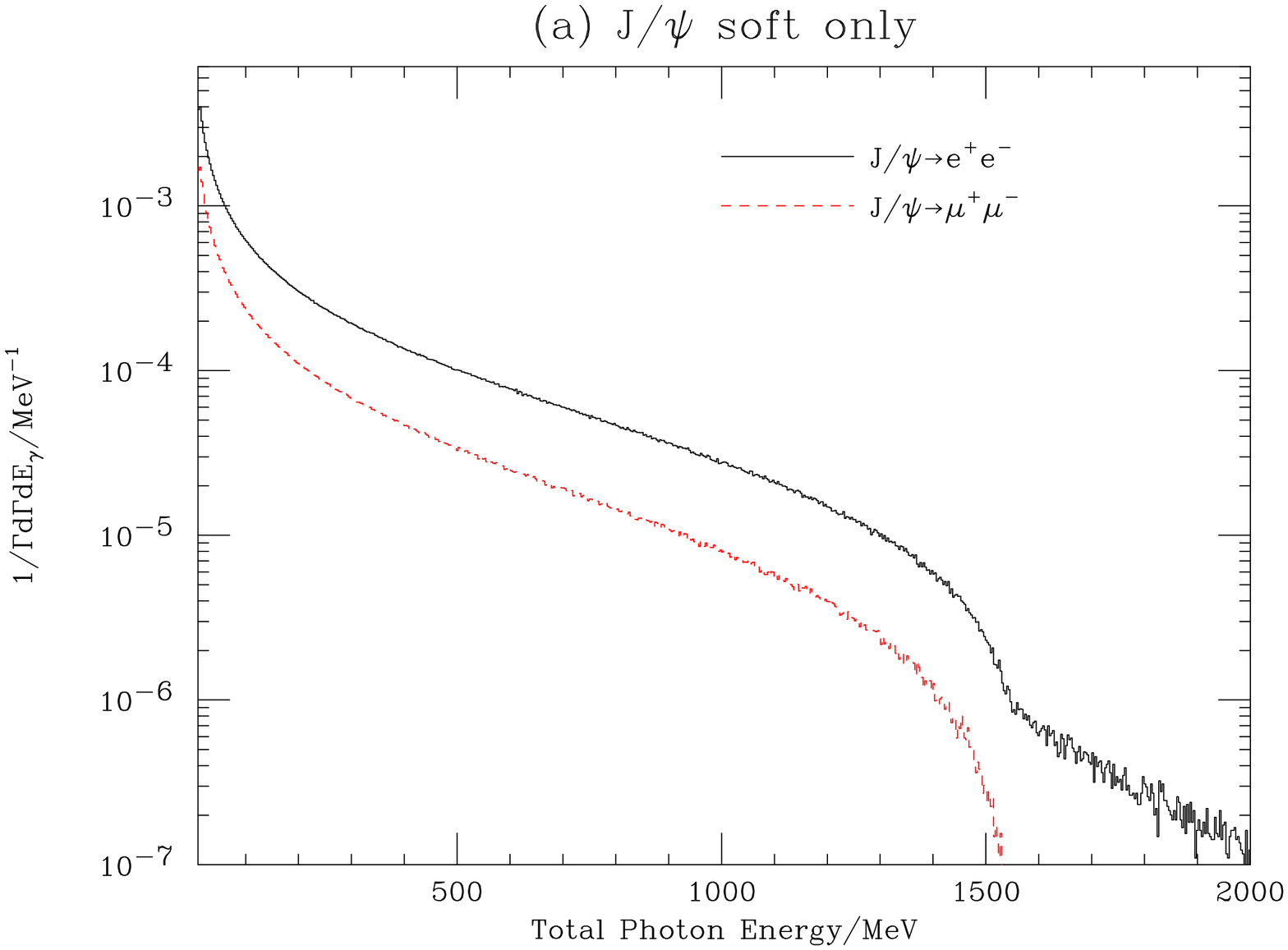}\hfill{}\includegraphics[%
  width=0.34\textwidth,
  angle=90]{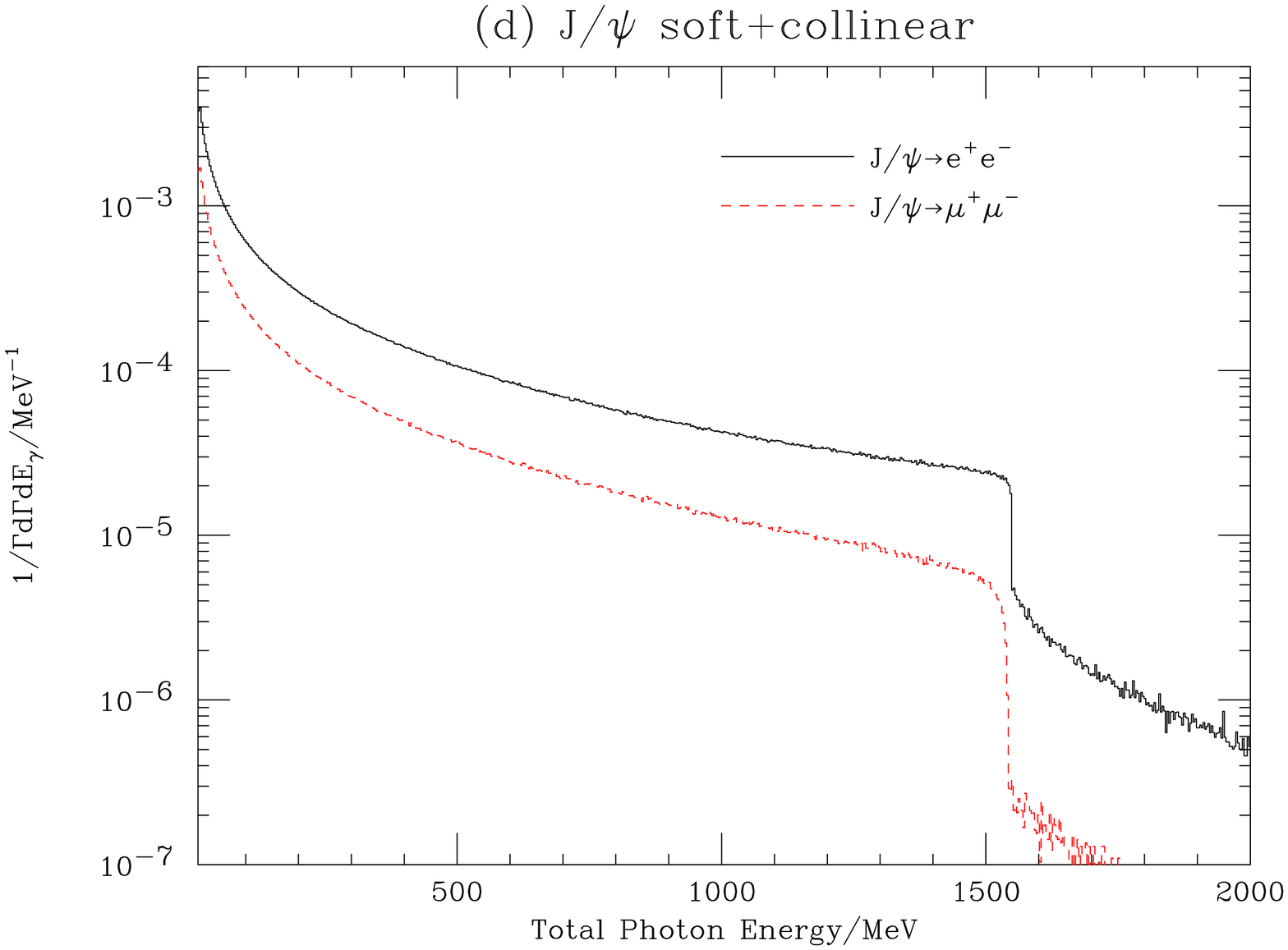}\\
\includegraphics[%
  width=0.34\textwidth,
  angle=90]{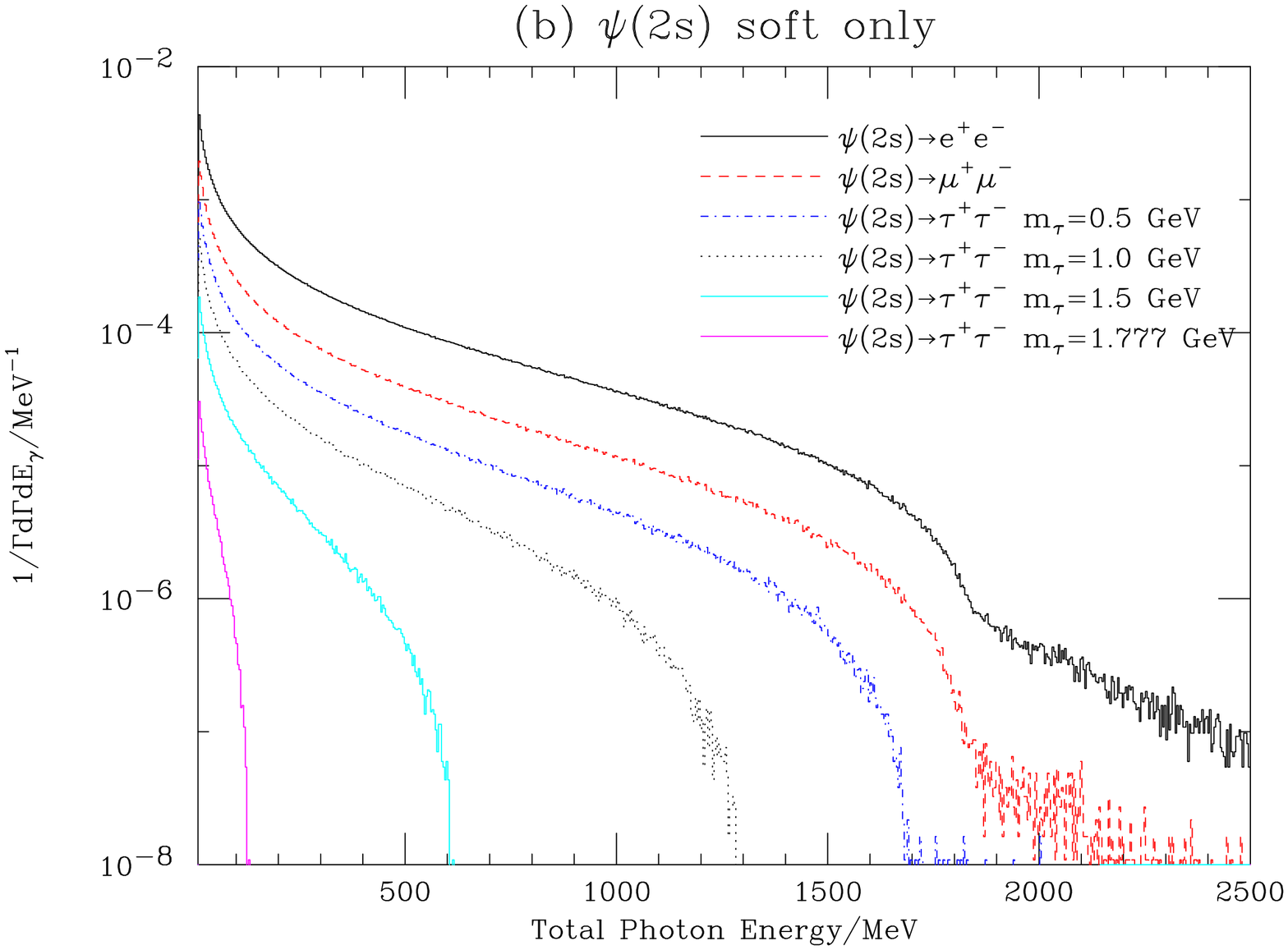}\hfill{}\includegraphics[%
  width=0.34\textwidth,
  angle=90]{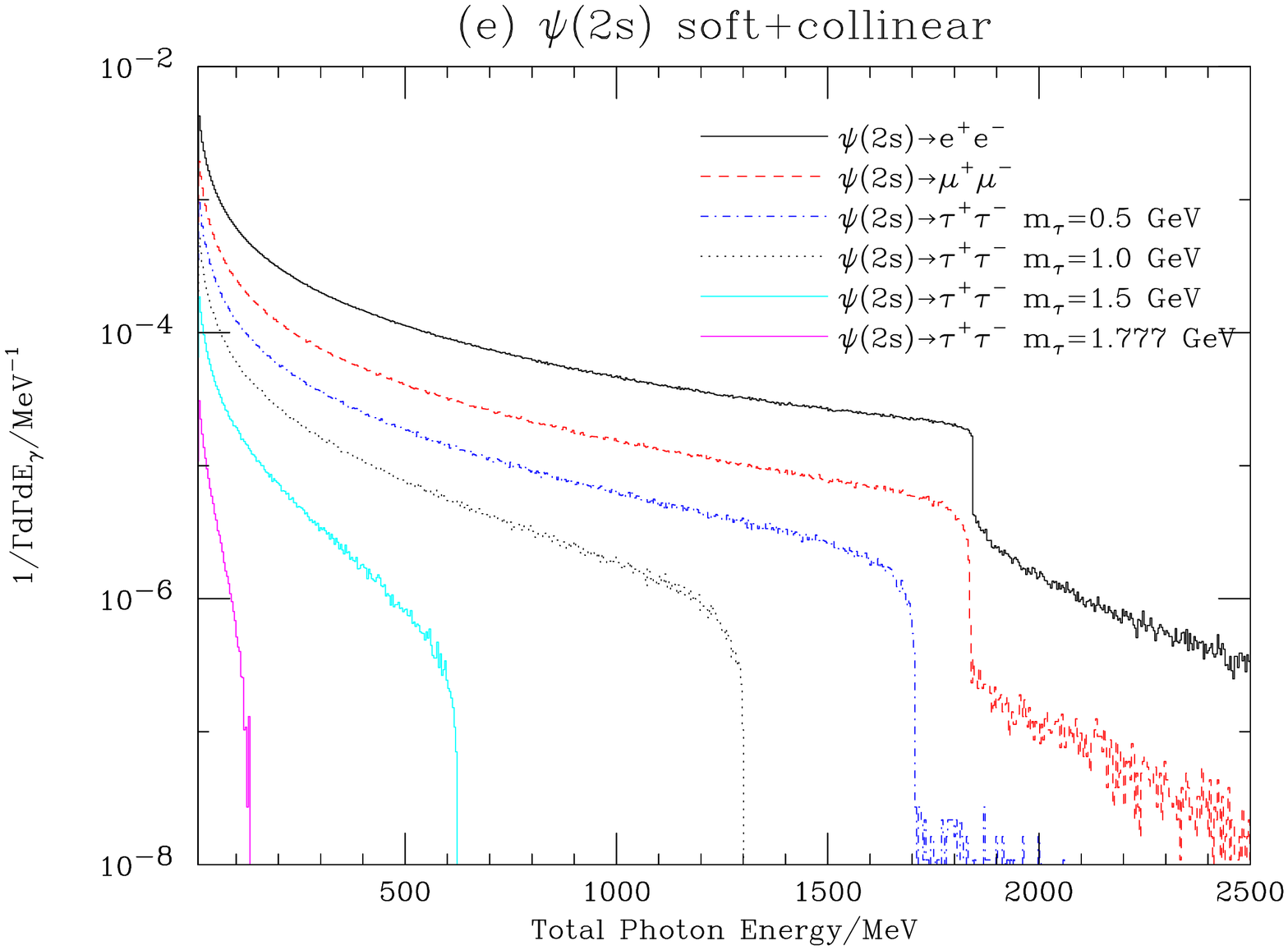}\\
\includegraphics[%
  width=0.34\textwidth,
  angle=90]{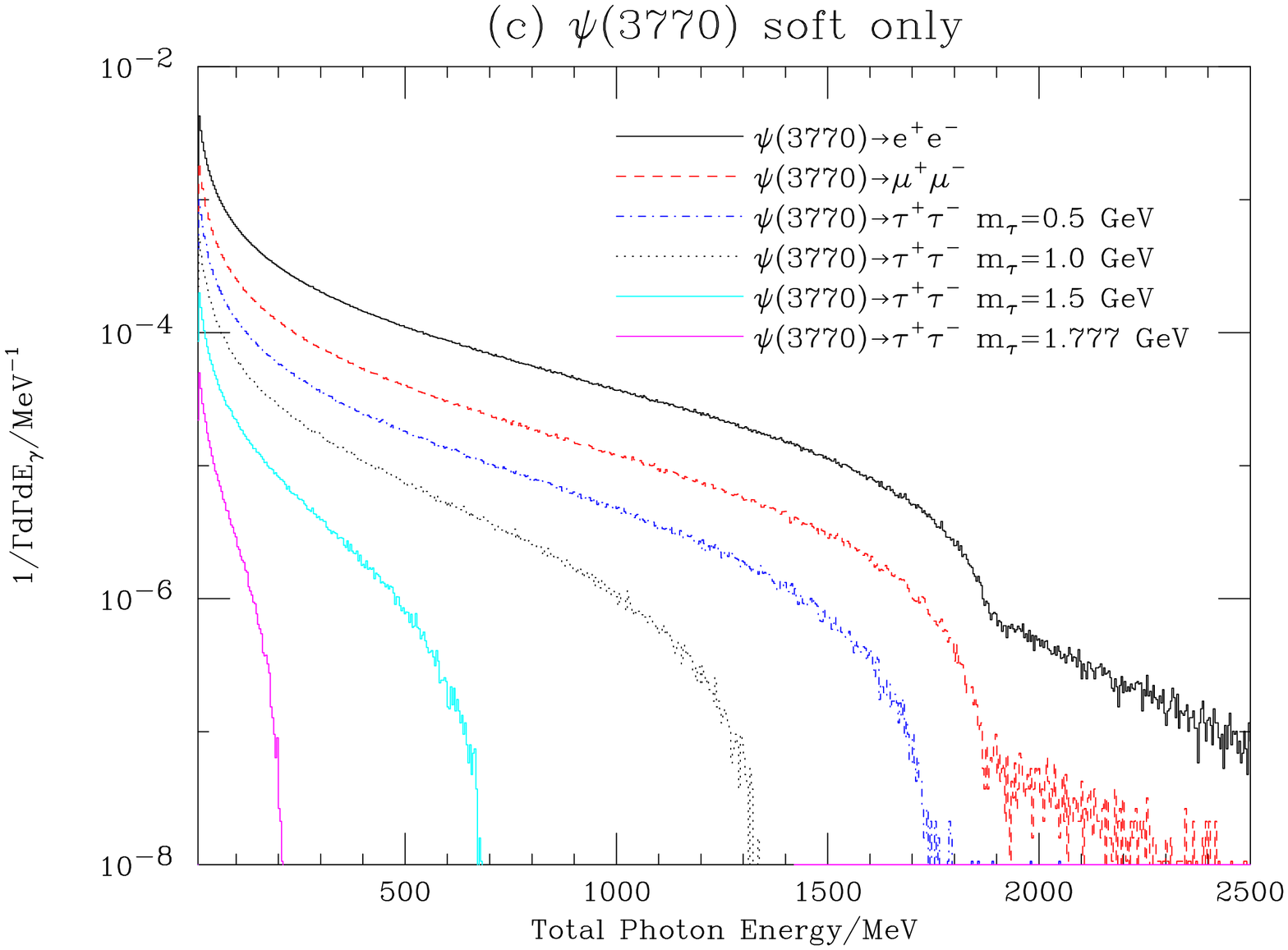}\hfill{}\includegraphics[%
  width=0.34\textwidth,
  angle=90]{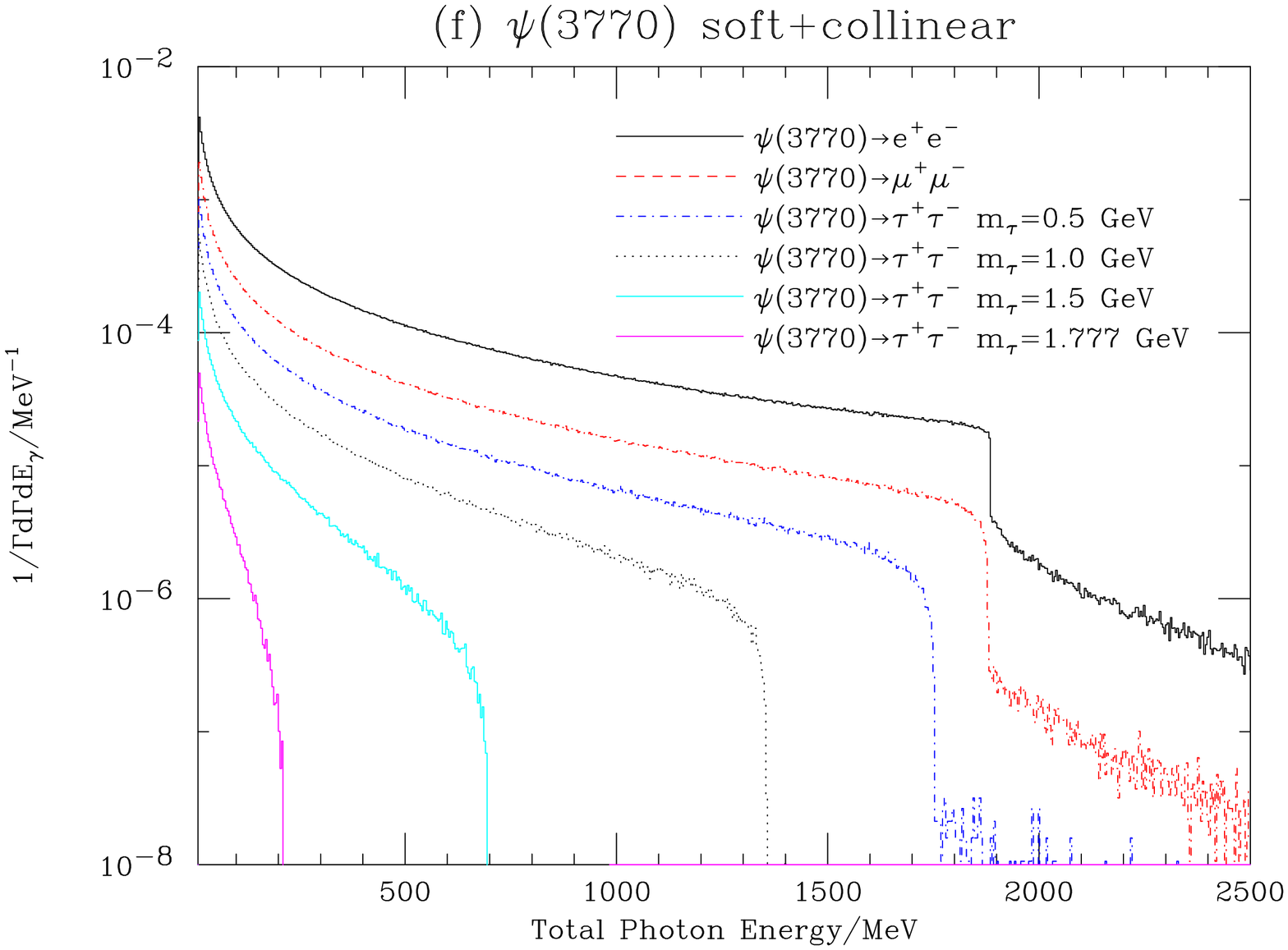}\\
\end{center}

\caption{\label{fig:charm}The total energy $\left(K_{0}\right)$ of the photons
radiated in leptonic decays of charmonium resonances is shown above
for ${\textrm{J}}/\psi$ (a/d), $\psi(2s)$ (b/e) and $\psi(3770)$
(c/f). The distributions on the left, figures (a), (b) and (c), are
obtained by truncating the infrared residuals at $\mathcal{O}\left(1\right)$,
whereas in (d), (e) and (f), the dipole splitting functions are used
to include the effects of hard collinear photons. In addition to the
real charged leptons we have included the effect of varying the $\tau$
mass to illustrate the mass dependence of the results. }
\end{figure}

\begin{figure}[!h]
\begin{center}\includegraphics[%
  width=0.34\textwidth,
  angle=90]{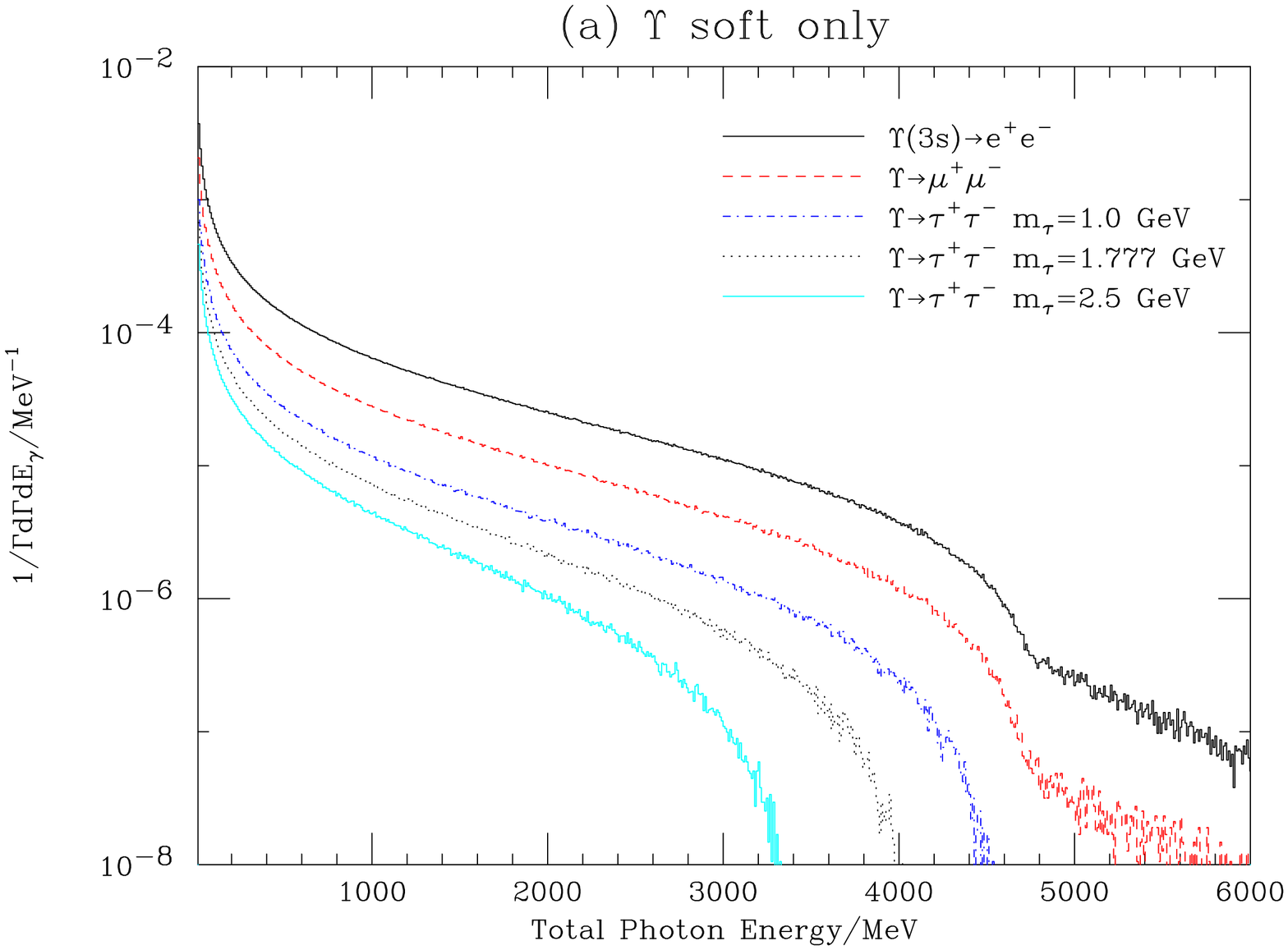}\hfill{}\includegraphics[%
  width=0.34\textwidth,
  angle=90]{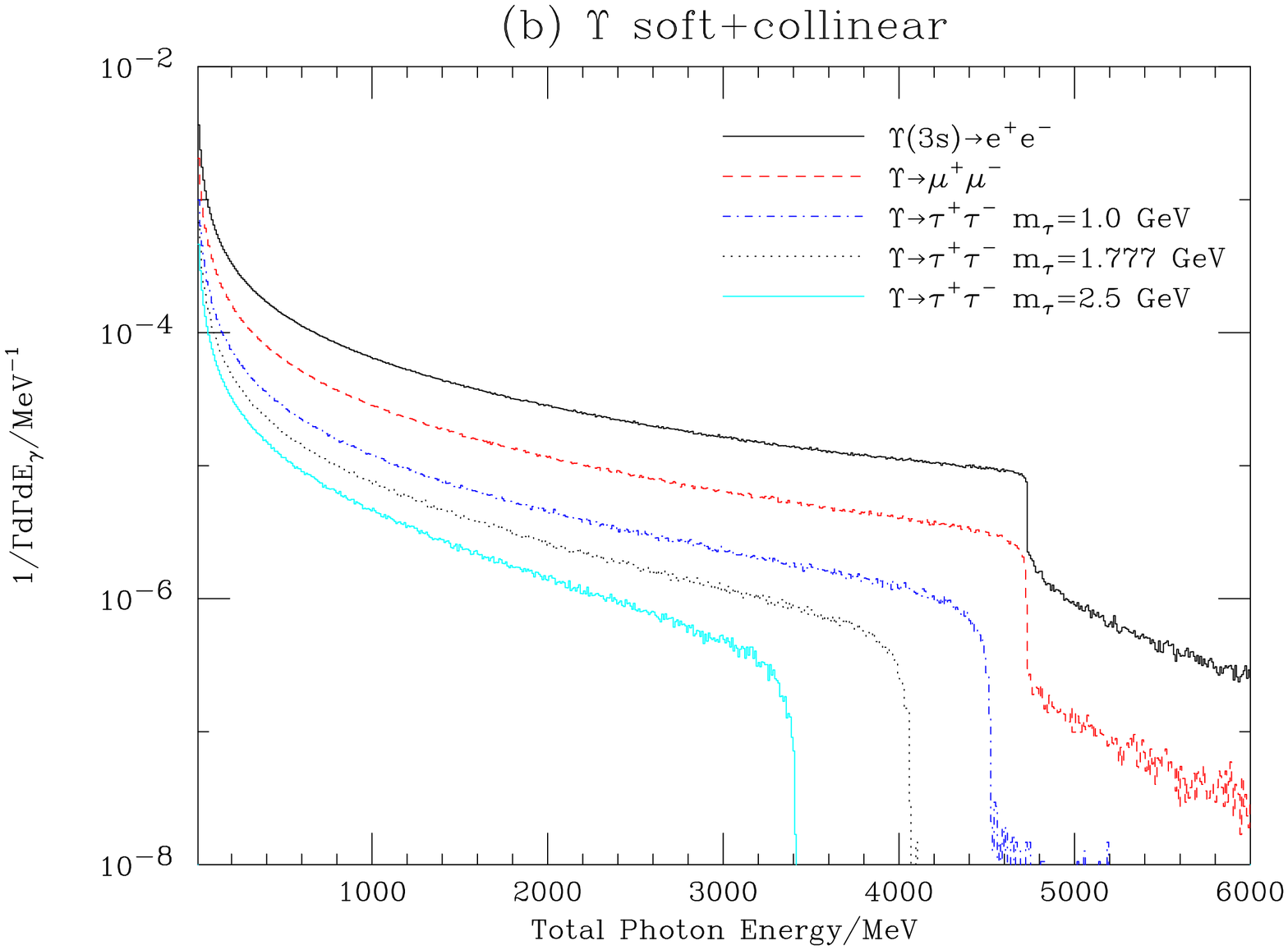}\end{center}

\caption{\label{fig:bottom}The total energy $\left(K_{0}\right)$ of the
photons radiated in leptonic decays of the bottomonium resonance $\Upsilon$.
The distribution on the left (a) is obtained by truncating the infrared
residuals at $\mathcal{O}\left(1\right)$, whereas in (b), the dipole
splitting functions $\left(\mathcal{D}_{ij}\right)$ are used to include
the effects of hard collinear emissions.}
\end{figure}

For the leptonic decays of the charmonium resonances and the $\Upsilon$
resonance, the total energy spectrum of the radiated photons is shown
in figures \ref{fig:charm} and \ref{fig:bottom} respectively. As
for $\mathrm{Z}$ decays, the effects of varying the $\tau$ mass
are included to show the mass dependence of the results. In the charmonium
decays to $\tau^{+}\tau^{-}$ pairs we see a suppression of QED radiation
since these decay modes are near the production threshold; for the
physical $\tau$ mass, this decay mode is not accessible in $\textrm{J}/\psi$
decays, while for $\psi\left(2s\right)$ and $\psi\left(3770\right)$
it is just below the threshold. The $\Upsilon$ resonance is significantly
more massive and therefore the associated photon energy spectrum more
closely resembles that seen for the case of the $\textrm{Z}$ boson.

These results show that the approach can be successfully applied to
both perturbative decays and non-perturbative hadronic decays.

\section{Conclusions\label{sec:Conclusions}}

In this paper we have presented a universal theoretical framework
for calculating QED radiative corrections to particle decays based
on the YFS formalism \cite{Yennie:1961ad} and the methods of \cite{Ward:1987jg,Bonneau-Martin,YFS2}.
The essence of this approach is a reorganization of the perturbation
series to resum all soft divergent QED logarithms. This formalism
led to the master formula presented in (\ref{eq:2.2}).

The master formula forms the basis of the Monte Carlo event generator
\textsf{SOPHTY}, which provides QED radiative corrections for decays
inside the \textsf{HERWIG++} generator. The Monte Carlo simulation
takes into account large soft photon logarithms to all orders. In
addition, the leading collinear logarithms are included to $\mathcal{O}\left(\alpha\right)$
by using the so-called dipole splitting functions and inferring the
associated virtual corrections with the aid of the Bloch-Nordsieck
and the KLN theorems.

Algorithms for the two basic {}``building-block'' cases of dipoles
comprising either two final-state particles, or the initial-state
particle and one of its decay products, are presented in section \ref{sec:Master-Equation}.
Although integrals like that of the master equation can generally
be readily performed with conventional Monte Carlo methods to give
weighted events, the manipulations required in order to produce unweighted
events with good efficiency (\emph{i.e.} an event generator) are non-trivial.

In designing these algorithms we were constrained by the requirement
that the QED radiation should be generated, as far as possible, independently
of the details of the main \textsf{HERWIG++} program, which should
provide the initial distribution of decay products. This was made
possible due to the form of the master equation, the universal nature
of the radiative corrections involved and our simplified crude distribution
from which we initially generate the photons. Care was taken to design
the algorithms to keep event weights as close to one as possible and
to avoid numerical instabilities. Key to realising these features
are importance sampling techniques and, importantly, a careful choice
of frame in which to generate the radiation.

Our algorithm was tested successfully for several different types
of particle decay produced by the \textsf{HERWIG++} generator. In
section \ref{sec:Results} we have shown results for the important
cases of $\mathrm{W}$ and $\mathrm{Z}$ decays to various lepton
species. In all cases the distributions show a smooth and stable behaviour
agreeing with our expectations. A preliminary comparison of the total
photon energy spectrum from the \textsf{WINHAC} generator shows good
agreement and provides a good check of our methods. The application
of the program to both hadronic and leptonic meson decays was also
illustrated in section \ref{sec:Results}.

There are several possible extensions of this work, for instance,
there are a number phenomenologically important decays for which the
full $\mathcal{O}\left(\alpha\right)$ corrections are known \emph{e.g.}
$\mathrm{W}$ and $\mathrm{Z}$ decays. As the code is designed to
readily allow these corrections to be included they will be implemented
in the near future. In addition, there are a small number of cases
inside the \textsf{HERWIG++} where we have to simulate multi~(\textit{i.e.}
greater than two) body decays and the extension of the algorithm to
these cases would be useful. This is the first use of the YFS approach
within the \textsf{HERWIG++} program and there are a number of other
potential applications, for example initial-state radiation in lepton
collisions, which may be pursued in the future.

In conclusion, we have applied the YFS formalism to the simulation
of QED radiation in particle decays. The simulation, \textsf{SOPHTY},
based on the results of this work can simulate QED radiation in a
wide range of particle decays and will be available in the next version
of the \textsf{HERWIG++} program.

\section*{Acknowledgements}

We would like to thank our collaborators on the \textsf{HERWIG++}
project for many fruitful discussions. This work was supported by
PPARC.

\appendix

\section{YFS Form Factors}

In this appendix we give the expressions we calculate for the YFS
form factors. For both the final-final dipole form factor and the
initial-final dipole form factor we use, $\beta_{i}=\left|\vec{p}_{i}\right|/E_{i}$
, to denote the velocity of particle $i$. Furthermore, in each case
we have assumed that the momenta obey $p=p_{1}+p_{2}$.

\subsection{Final-Final Dipoles\label{sub:YFS_FF_Form_Factor}}

Below in (\ref{eq:7.1.1}) we have the YFS form factor for a pair
of final state charged particles with momenta $p_{1}$, $p_{2}$ whose
combined momentum is $p$. Expression (\ref{eq:7.1.1}) is valid in
the frame $\vec{p}_{1}=-\vec{p}_{2}$.

\begin{equation}
\begin{array}{lcl}
Y\left(p_{1},p_{2},\Omega\right) & = & -\frac{\alpha}{2\pi}\textrm{ }Z_{1}Z_{2}\textrm{ }\hat{Y}\left(p_{1},p_{2},\Omega\right)\\
\hat{Y}\left(p_{1},p_{2},\Omega\right) & = & \left(4-2\left(\frac{1+\beta_{1}\beta_{2}}{\beta_{1}+\beta_{2}}\right)\left(\ln\left(\frac{1+\beta_{1}}{1-\beta_{1}}\right)+\ln\left(\frac{1+\beta_{2}}{1-\beta_{2}}\right)\right)\right)\ln\left(\frac{\sqrt{p^{2}}}{2\omega}\right)\\
 & - & \frac{1}{2}\ln\left(\frac{p^{2}}{p_{1}^{2}}\right)-\frac{1}{2}\ln\left(\frac{p^{2}}{p_{2}^{2}}\right)-2\\
 & + & \left(\frac{\beta_{2}-\beta_{1}\beta_{2}}{\beta_{1}+\beta_{2}}\right)\ln\left(\frac{\beta_{2}-\beta_{1}\beta_{2}}{\beta_{1}+\beta_{2}}\right)+\left(\frac{\beta_{1}+\beta_{1}\beta_{2}}{\beta_{1}+\beta_{2}}\right)\ln\left(\frac{\beta_{1}+\beta_{1}\beta_{2}}{\beta_{1}+\beta_{2}}\right)\\
 & + & \left(\frac{\beta_{1}-\beta_{1}\beta_{2}}{\beta_{1}+\beta_{2}}\right)\ln\left(\frac{\beta_{1}-\beta_{1}\beta_{2}}{\beta_{1}+\beta_{2}}\right)+\left(\frac{\beta_{2}+\beta_{1}\beta_{2}}{\beta_{1}+\beta_{2}}\right)\ln\left(\frac{\beta_{2}+\beta_{1}\beta_{2}}{\beta_{1}+\beta_{2}}\right)\\
 & + & \frac{1+\beta_{1}\beta_{2}}{\beta_{1}+\beta_{2}}\left(\frac{1}{2}\ln^{2}\left(\frac{\beta_{2}-\beta_{1}\beta_{2}}{\beta_{1}+\beta_{2}}\right)-\frac{1}{2}\ln^{2}\left(\frac{\beta_{2}+\beta_{1}\beta_{2}}{\beta_{1}+\beta_{2}}\right)\right)\\
 & + & \frac{1+\beta_{1}\beta_{2}}{\beta_{1}+\beta_{2}}\left(\frac{1}{2}\ln^{2}\left(\frac{\beta_{1}-\beta_{1}\beta_{2}}{\beta_{1}+\beta_{2}}\right)-\frac{1}{2}\ln^{2}\left(\frac{\beta_{1}+\beta_{1}\beta_{2}}{\beta_{1}+\beta_{2}}\right)\right)\\
 & - & \frac{1+\beta_{1}\beta_{2}}{\beta_{1}+\beta_{2}}\left(2\textrm{Li}_{2}\left(-\frac{1-\beta_{1}}{2\beta_{1}}\right)+2\textrm{Li}_{2}\left(-\frac{1-\beta_{2}}{2\beta_{2}}\right)\right)\\
 & - & \frac{1+\beta_{1}\beta_{2}}{\beta_{1}+\beta_{2}}\left(2\textrm{Li}_{2}\left(\frac{2\beta_{1}}{1+\beta_{1}}\right)+2\textrm{Li}_{2}\left(\frac{2\beta_{2}}{1+\beta_{2}}\right)\right)\\
 & - & \frac{1+\beta_{1}\beta_{2}}{\beta_{1}+\beta_{2}}\left(\ln\left(\frac{1-\beta_{1}}{2\beta_{1}}\right)\ln\left(\frac{1+\beta_{1}}{2\beta_{1}}\right)+\ln\left(\frac{1-\beta_{2}}{2\beta_{2}}\right)\ln\left(\frac{1+\beta_{2}}{2\beta_{2}}\right)\right)\\
 & + & \frac{1+\beta_{1}\beta_{2}}{\beta_{1}+\beta_{2}}\ln\left(\frac{2\beta_{1}\beta_{2}}{\beta_{1}+\beta_{2}}\right)\left(\ln\left(\frac{\beta_{2}-\beta_{1}\beta_{2}}{\beta_{1}+\beta_{2}}\right)-\ln\left(\frac{\beta_{2}+\beta_{1}\beta_{2}}{\beta_{1}+\beta_{2}}\right)\right)\\
 & + & \frac{1+\beta_{1}\beta_{2}}{\beta_{1}+\beta_{2}}\ln\left(\frac{2\beta_{1}\beta_{2}}{\beta_{1}+\beta_{2}}\right)\left(\ln\left(\frac{\beta_{1}-\beta_{1}\beta_{2}}{\beta_{1}+\beta_{2}}\right)-\ln\left(\frac{\beta_{1}+\beta_{1}\beta_{2}}{\beta_{1}+\beta_{2}}\right)\right)\\
 & + & \frac{1+\beta_{1}\beta_{2}}{\beta_{1}+\beta_{2}}\left(\frac{4\pi^{2}}{3}\right)-\frac{1}{\beta_{1}}\ln\left(\frac{1-\beta_{1}}{1+\beta_{1}}\right)-\frac{1}{\beta_{2}}\ln\left(\frac{1-\beta_{2}}{1+\beta_{2}}\right)\\
 & - & \frac{1+\beta_{1}\beta_{2}}{\beta_{1}+\beta_{2}}\left(\frac{1}{2}\ln^{2}\left(\frac{1-\beta_{1}}{1+\beta_{1}}\right)+\frac{1}{2}\ln^{2}\left(\frac{1-\beta_{2}}{1+\beta_{2}}\right)\right)\end{array}\label{eq:7.1.1}\end{equation}
We have checked that for the case $\beta_{1}=\beta_{2}$, in the limit
$\beta_{1}\rightarrow1$, expression (\ref{eq:7.1.1}) reduces to
\begin{equation}
Y\left(p_{1},p_{2},\Omega\right)=-\frac{\alpha}{\pi}\textrm{ }Z_{1}Z_{2}\textrm{ }\left(2\ln\left(\frac{2\omega}{\sqrt{p^{2}}}\right)\left(\ln\left(\frac{p^{2}}{p_{1}^{2}}\right)-1\right)+\frac{1}{2}\ln\left(\frac{p^{2}}{p_{1}^{2}}\right)-1+\frac{\pi^{2}}{3}\right),\label{eq:7.1.2}\end{equation}
 in agreement with the previously obtained results \cite{Jadach:2000ir}.

\subsection{Initial-Final Dipoles\label{sub:YFS_IF_Form_Factor}}

In equation (\ref{eq:7.2.1}) we have the YFS form factor for a pair
of charged particles of momentum $p$ and $p_{1}$, $\left(p_{2}=p-p_{1}\right)$,
evaluated in the rest frame of $p$.\begin{equation}
\begin{array}{lcl}
Y\left(p,p_{1},\Omega\right) & = & \frac{\alpha}{2\pi}\textrm{ }Z_{p}Z_{1}\textrm{ }\hat{Y}\left(p,p_{1},\Omega\right)\\
\hat{Y}\left(p,p_{1},\Omega\right) & = & \ln\left(\frac{p^{2}}{4\omega^{2}}\right)+\ln\left(\frac{p_{1}^{2}}{4\omega^{2}}\right)-\frac{1}{\beta_{1}}\ln\left(\frac{1+\beta_{1}}{1-\beta_{1}}\right)\ln\left(\frac{p_{2}^{2}}{4\omega^{2}}\right)\\
 & + & \frac{1}{2}\ln\left(\frac{p_{2}^{4}}{p^{2}p_{1}^{2}}\right)-\frac{1}{\beta_{1}}\ln\left(\frac{1-\beta_{1}}{1+\beta_{1}}\right)-\frac{1}{2\beta_{1}}\ln^{2}\left(\frac{1-\beta_{1}}{1+\beta_{1}}\right)\\
 & + & \left(\frac{\beta_{1}+\beta_{2}}{\beta_{1}-\beta_{1}\beta_{2}}\right)\ln\left(\frac{\beta_{1}+\beta_{2}}{\beta_{1}-\beta_{1}\beta_{2}}\right)-\left(\frac{\beta_{2}+\beta_{1}\beta_{2}}{\beta_{1}-\beta_{1}\beta_{2}}\right)\ln\left(\frac{\beta_{2}+\beta_{1}\beta_{2}}{\beta_{1}-\beta_{1}\beta_{2}}\right)\\
 & + & \left(\frac{\beta_{1}+\beta_{2}}{\beta_{1}+\beta_{1}\beta_{2}}\right)\ln\left(\frac{\beta_{1}+\beta_{2}}{\beta_{1}+\beta_{1}\beta_{2}}\right)-\left(\frac{\beta_{2}-\beta_{1}\beta_{2}}{\beta_{1}+\beta_{1}\beta_{2}}\right)\ln\left(\frac{\beta_{2}-\beta_{1}\beta_{2}}{\beta_{1}+\beta_{1}\beta_{2}}\right)\\
 & + & \frac{1}{2\beta_{1}}\ln^{2}\left(\frac{\beta_{1}+\beta_{2}}{\beta_{1}-\beta_{1}\beta_{2}}\right)-\frac{1}{2\beta_{1}}\ln^{2}\left(\frac{\beta_{2}+\beta_{1}\beta_{2}}{\beta_{1}-\beta_{1}\beta_{2}}\right)\\
 & + & \frac{1}{2\beta_{1}}\ln^{2}\left(\frac{\beta_{2}-\beta_{1}\beta_{2}}{\beta_{1}+\beta_{1}\beta_{2}}\right)-\frac{1}{2\beta_{1}}\ln^{2}\left(\frac{\beta_{1}+\beta_{2}}{\beta_{1}+\beta_{1}\beta_{2}}\right)\\
 & + & \frac{2}{\beta_{1}}\textrm{Li}_{2}\left(-\frac{1-\beta_{2}}{2\beta_{2}}\right)-\frac{2}{\beta_{1}}\textrm{Li}_{2}\left(-\frac{\left(1-\beta_{1}\right)\left(1-\beta_{2}\right)}{2\left(\beta_{1}+\beta_{2}\right)}\right)-\frac{2}{\beta_{1}}\textrm{Li}_{2}\left(\frac{2\beta_{1}}{1+\beta_{1}}\right)\\
 & + & \frac{1}{\beta_{1}}\ln\left(\frac{\beta_{1}+\beta_{2}}{\beta_{1}+\beta_{1}\beta_{2}}\right)\ln\left(\frac{1+\beta_{2}}{2\beta_{2}}\right)-\frac{1}{\beta_{1}}\ln\left(\frac{\beta_{2}\left(1-\beta_{1}\right)}{\beta_{1}\left(1+\beta_{2}\right)}\right)\ln\left(\frac{\left(1+\beta_{1}\right)\left(1+\beta_{2}\right)}{2\left(\beta_{1}+\beta_{2}\right)}\right)\\
 & - & \frac{1}{\beta_{1}}\ln\left(\frac{2\beta_{2}}{\beta_{1}}\left(\frac{\beta_{1}+\beta_{2}}{1-\beta_{2}^{2}}\right)\right)\ln\left(\frac{\beta_{1}+\beta_{2}}{\beta_{2}-\beta_{1}\beta_{2}}\right).\end{array}\label{eq:7.2.1}\end{equation}
 We use $Z_{p}$ to denote the charge on the particle with momentum
$p$. For the special case that $\left(p-p_{2}\right)^{2}$ is zero,
as in leptonic $\mathrm{W}^{\pm}$ decay with a massless neutrino,
we use the a more specialized compact form, to avoid potential numerical
problems: \begin{equation}
\begin{array}{lcl}
Y\left(p,p_{1},\Omega\right) & = & \frac{\alpha}{2\pi}\textrm{ }Z_{p}Z_{1}\textrm{ }Y\left(p,p_{1},\Omega\right)\\
\hat{Y}\left(p,p_{1},\Omega\right) & = & \ln\left(\frac{p^{2}}{4\omega^{2}}\right)+\ln\left(\frac{p_{1}^{2}}{4\omega^{2}}\right)-\frac{1}{\beta_{1}}\ln\left(\frac{1+\beta_{1}}{1-\beta_{1}}\right)\ln\left(\frac{p^{2}-p_{1}^{2}}{4\omega^{2}}\right)-\frac{1}{2}\ln\left(\frac{1-\beta_{1}^{2}}{4\beta_{1}^{2}}\right)\\
 & + & \left(\frac{1+\beta_{1}}{2\beta_{1}}\right)\ln\left(\frac{1+\beta_{1}}{2\beta_{1}}\right)-\left(\frac{1-\beta_{1}}{2\beta_{1}}\right)\ln\left(\frac{1-\beta_{1}}{2\beta_{1}}\right)-\frac{1}{\beta_{1}}\ln\left(\frac{1-\beta_{1}}{1+\beta_{1}}\right)+1\\
 & + & \frac{1}{2\beta_{1}}\ln^{2}\left(\frac{1-\beta_{1}}{2\beta_{1}}\right)-\frac{1}{2\beta_{1}}\ln^{2}\left(\frac{1+\beta_{1}}{2\beta_{1}}\right)-\frac{1}{2\beta_{1}}\ln^{2}\left(\frac{1-\beta_{1}}{1+\beta_{1}}\right)-\frac{2}{\beta_{1}}{\textrm{{Li}}}_{2}\left(\frac{2\beta_{1}}{1+\beta_{1}}\right).\end{array}\label{eq:7.2.2}\end{equation}
 We have checked, analytically, that in the limit $\beta_{2}\rightarrow1$
the virtual contributions $\left(\mathcal{R}e\textrm{ }B\left(p,p_{1}\right)\right)$
to $Y\left(p,p_{1},\Omega\right)$ inside (\ref{eq:7.2.1}) are equal
to those in (\ref{eq:7.2.2}). In both cases the real contributions
are identical, they do not involve $\beta_{2}$, naturally as these
contributions should only involve the moving charge in the dipole.
Finally, as a check we observe that, dropping terms smaller than $\mathcal{O}\left(p_{1}^{2}/p^{2}\right)$
the form factor (\ref{eq:7.2.2}) agrees exactly with the corresponding
expression in \cite{Placzek:2003zg}.

\section{Generation of the Dipole Distributions}

In this appendix we describe how to generate the photon momenta from
the dipole radiation functions.

\subsection{Final-Final Dipoles\label{sub:Final-Final_Generation}}

Consider the integral of the dipole function in the rest frame of
$p_{1}+p_{2}$\begin{equation}
\int\frac{\mathrm{d}^{3}k}{k_{0}}\textrm{ }\tilde{S}\left(p_{1},p_{2},k\right)=\frac{\alpha}{4\pi^{2}}\textrm{ }Z_{1}Z_{2}\int\mathrm{dc}\textrm{ }\mathrm{d}\phi\textrm{ }\mathrm{d}k_{0}\textrm{ }k_{0}\left(\frac{p_{1}}{p_{1}\cdot k}-\frac{p_{2}}{p_{2}\cdot k}\right)^{2}.\label{eq:8.1.1}\end{equation}
 We choose to define the photon momenta as being with respect to $p_{1}$,
with $\mathrm{c}\equiv\cos\theta$ \emph{i.e.}\begin{equation}
\begin{array}{lcl}
p_{1}.k & = & E_{1}k_{0}\left(1-\beta_{1}\mathrm{c}\right),\\
p_{2}.k & = & E_{2}k_{0}\left(1+\beta_{2}\mathrm{c}\right).\end{array}\label{eq:8.1.2}\end{equation}
 Using this representation of the momenta the integral can be rewritten
as \begin{equation}
\begin{array}{lcl}
\int\frac{\mathrm{d}^{3}k}{k^{0}}\textrm{ }\tilde{S}\left(p_{1},p_{2},k\right) & = & -\frac{\alpha}{4\pi^{2}}\textrm{ }Z_{1}Z_{2}\int\mathrm{dc}\textrm{ }\mathrm{d}\phi\textrm{ }\mathrm{d}\ln k_{0}\left(-\frac{1-\beta_{1}^{2}}{\left(1-\beta_{1}\mathrm{c}\right)^{2}}+\frac{2\left(1+\beta_{1}\beta_{2}\right)}{\left(1-\beta_{1}\mathrm{c}\right)\left(1+\beta_{2}\mathrm{c}\right)}-\frac{1-\beta_{2}^{2}}{\left(1+\beta_{2}\mathrm{c}\right)^{2}}\right).\end{array}\label{eq:8.1.3}\end{equation}
 The photon momenta can be generated according to this distribution
in the following way. 

\begin{enumerate}
\item First the magnitude of the photon's momentum is generated logarithmically
between $\omega$, the minimum photon energy cut-off, and the maximum
possible photon energy, $E_{\textrm{max}}$, \emph{i.e.}\begin{equation}
k_{0}=\omega\left(\frac{E_{\textrm{max}}}{\omega}\right)^{\mathcal{R}},\label{eq:8.1.4}\end{equation}
 where $\mathcal{R}$ is a random number uniformly distributed between
0~and~1. As the photon momenta are generated in the rest frame of
the dipole the maximum energy of the photon is \begin{equation}
E_{\textrm{max}}=\frac{M}{2}\left(\frac{M}{m_{1}+m_{2}}-\frac{m_{1}+m_{2}}{M}\right).\label{eq:8.1.5}\end{equation}

\item The azimuthal angle $\phi$ is randomly generated between $0$~and~$2\pi$. 
\item The generation of the polar angle $\theta$ is more complicated. The
polar angle is generated by only using the interference term \emph{i.e.}
neglecting mass terms. This term is first rewritten \begin{equation}
\frac{1}{\left(1-\beta_{1}\mathrm{c}\right)\left(1+\beta_{2}\mathrm{c}\right)}=\frac{\beta_{1}\beta_{2}}{\beta_{1}+\beta_{2}}\left(\frac{1}{\beta_{2}\left(1-\beta_{1}\mathrm{c}\right)}+\frac{1}{\beta_{1}\left(1+\beta_{2}\mathrm{c}\right)}\right).\label{eq:8.1.6}\end{equation}
 The angle can then be generated according to this distribution by
generating the angle according to the distribution $\left(1-\beta_{1}\mathrm{c}\right)^{-1}$
with probability \begin{equation}
P_{1}=\frac{\ln\left(\frac{1+\beta_{1}}{1-\beta_{1}}\right)}{\ln\left(\frac{1+\beta_{1}}{1-\beta_{1}}\right)+\ln\left(\frac{1+\beta_{2}}{1-\beta_{2}}\right)}\label{eq:8.1.7}\end{equation}
 and according to the distribution $\left(1+\beta_{2}\mathrm{c}\right)^{-1}$
with probability $P_{2}=1-P_{1}$.

The full distribution can easily be generated using rejection techniques,
the rejection weight \begin{equation}
\mathcal{W}=\frac{\left(-\frac{1-\beta_{1}^{2}}{\left(1-\beta_{1}\mathrm{c}\right)^{2}}+\frac{2\left(1+\beta_{1}\beta_{2}\right)}{\left(1-\beta_{1}\mathrm{c}\right)\left(1+\beta_{2}\mathrm{c}\right)}-\frac{1-\beta_{2}^{2}}{\left(1+\beta_{2}\mathrm{c}\right)^{2}}\right)}{\frac{2\left(1+\beta_{1}\beta_{2}\right)}{\left(1-\beta_{1}\mathrm{c}\right)\left(1+\beta_{2}\mathrm{c}\right)}}\leq1\label{eq:8.1.8}\end{equation}
 is less than one.

In practice we sometimes choose not to generate the angle according
to the full distribution initially \emph{i.e.} we postpone the latter
rejection step until the event is generated in full. This is because
the inclusion of the mass terms leads to a depletion of radiation
in the direction of the charged particles, the {}``dead-cone'' \cite{Marchesini:1989yk}.
However this dead-cone can be filled by hard radiation and if the
$\bar{\beta}_{1}^{1}$ corrections are included. In this case, if
the generation of the angles is done according to the full distribution,
the algorithm becomes very inefficient.

\end{enumerate}
In order to calculate the crude distribution we require the average
photon multiplicity, which is given by the integral of the dipole
function\begin{equation}
\begin{array}{lll}
 & \int\frac{\mathrm{d}^{3}k}{k_{0}}\textrm{ }\tilde{S}\left(p_{1},p_{2},k\right)\\
= & -\frac{\alpha}{\pi}\textrm{ }Z_{1}Z_{2}\ln\left(\frac{E_{max}}{\omega}\right)\left(\left(\frac{1+\beta_{1}\beta_{2}}{\beta_{1}+\beta_{2}}\right)\ln\left(\frac{\left(1-\beta_{1}\right)\left(1-\beta_{2}\right)}{\left(1+\beta_{1}\right)\left(1+\beta_{2}\right)}\right)-2\right) & \mathrm{\textrm{full distribution,}}\\
= & -\frac{\alpha}{\pi}\textrm{ }Z_{1}Z_{2}\left(\frac{1+\beta_{1}\beta_{2}}{\beta_{1}+\beta_{2}}\right)\ln\left(\frac{E_{max}}{\omega}\right)\ln\left(\frac{\left(1-\beta_{1}\right)\left(1-\beta_{2}\right)}{\left(1+\beta_{1}\right)\left(1+\beta_{2}\right)}\right) & \textrm{neglecting mass terms.}\end{array}\label{eq:8.1.9}\end{equation}

\subsection{Initial-Final Dipoles\label{sub:Initial-Final_Generation}}

Consider the integral of the dipole function in the rest frame of
$p$\begin{equation}
\int\frac{\mathrm{d}^{3}k}{k_{0}}\textrm{ }\tilde{S}\left(p,p_{1},k\right)=-\frac{\alpha}{4\pi^{2}}\textrm{ }Z_{p}Z_{1}\int\mathrm{dc}\textrm{ }\mathrm{d}\phi\textrm{ }\mathrm{d}k_{0}\textrm{ }k_{0}\left(\frac{p}{p\cdot k}-\frac{p_{1}}{p_{1}\cdot k}\right)^{2},\label{eq:8.2.1}\end{equation}
 where $Z_{p}$ denotes the charge on the decaying particle. As before
we choose to define the photon momenta as being with respect to $p_{1}$
hence the integral may be rewritten \begin{equation}
\int\frac{\mathrm{d}^{3}k}{k_{0}}\textrm{ }\tilde{S}\left(p,p_{1},k\right)=\frac{\alpha}{4\pi^{2}}\textrm{ }Z_{p}Z_{1}\int\mathrm{dc}\textrm{ }\mathrm{d}\phi\textrm{ }\mathrm{d}\ln k_{0}\textrm{ }\frac{\beta_{1}^{2}\left(1-\mathrm{c}^{2}\right)}{\left(1-\beta_{1}\mathrm{c}\right)^{2}}.\label{eq:8.2.2}\end{equation}
 The photon energy and azimuthal angle are generated in exactly the
same way as in appendix (\ref{sub:Final-Final_Generation}) with the
exception that now, to guarantee the possibility of conserving momentum
in the decaying particle's rest frame, the maximum allowable energy
of the photons is \begin{equation}
E_{\textrm{max}}=\frac{M}{2}\left(1-\frac{\left(m_{1}+m_{2}\right)^{2}}{M^{2}}\right).\label{eq:8.2.3}\end{equation}

The generation of the polar angle $\theta$ is more straightforward
than for the final-final dipole. Omitting the mass terms $p_{1}^{2}$
and $p^{2}$ in (\ref{eq:8.2.1}) leads to the replacement

\begin{equation}
\textrm{ }\frac{\beta_{1}^{2}\left(1-\mathrm{c}^{2}\right)}{\left(1-\beta_{1}\mathrm{c}\right)^{2}}\rightarrow\frac{2}{1-\beta_{1}\mathrm{c}},\label{eq:8.2.4}\end{equation}
 which may be generated by the simple mapping\begin{equation}
\mathrm{c}=\frac{1}{\beta_{1}}\left(1-\left(1+\beta_{1}\right)\left(\frac{1-\beta_{1}}{1+\beta_{1}}\right)^{\mathcal{R}}\right).\label{eq:8.2.5}\end{equation}
 The full distribution can be recovered by weighting and rejecting
the events from this approximate distribution, with weight \begin{equation}
\mathcal{W}=\frac{\beta_{1}^{2}\left(1-\mathrm{c}^{2}\right)}{2\left(1-\beta_{1}\mathrm{c}\right)}\leq1.\label{eq:8.2.6}\end{equation}

As with the final-final dipole the integral of the dipole function
gives the average photon multiplicity for $\Gamma_{\mathrm{crude}}$:\begin{equation}
\begin{array}{lll}
 & \int\frac{\mathrm{d}^{3}k}{k_{0}}\textrm{ }\tilde{S}\left(p,p_{1},k\right)\\
= & \frac{\alpha}{\pi}\textrm{ }Z_{p}Z_{1}\ln\left(\frac{E_{\mathrm{max}}}{\omega}\right)\left(\frac{1}{\beta_{1}}\ln\left(\frac{1+\beta_{1}}{1-\beta_{1}}\right)-2\right) & \mathrm{\textrm{full distribution}}\\
= & \frac{\alpha}{\pi}\textrm{ }Z_{p}Z_{1}\ln\left(\frac{E_{\mathrm{max}}}{\omega}\right)\left(\frac{1}{\beta_{1}}\ln\left(\frac{1+\beta_{1}}{1-\beta_{1}}\right)\right) & \textrm{neglecting mass terms.}\end{array}\label{eq:8.2.7}\end{equation}

\bibliographystyle{/usr/share/texmf/tex/jhep/JHEP}
\bibliography{YFS}

\end{document}